\documentclass{article}
\usepackage{PRIMEarxiv}

\usepackage[utf8]{inputenc} 
\usepackage[T1]{fontenc}    
\usepackage{hyperref}       
\usepackage{url}            
\usepackage{booktabs}       
\usepackage{amsmath, amsthm, amsfonts}       
\usepackage{nicefrac}       
\usepackage{microtype}      
\usepackage{fancyhdr}       
\usepackage{graphicx}       
\graphicspath{{media/}}     
\usepackage{balance} 
\usepackage{algorithm}
\usepackage{algpseudocode}
\usepackage{subcaption}
\usepackage{paralist}
\usepackage{appendix}
\usepackage{float}
\usepackage{latexsym}
\usepackage{amssymb}
\usepackage{booktabs}
\usepackage{enumitem}
\usepackage{color}
\usepackage{algorithm}
\usepackage{algpseudocode}
\usepackage{subcaption}
\usepackage{amsmath, amsthm, amsfonts}
\usepackage{paralist}
\usepackage{appendix}
\usepackage{float}
\usepackage{bbold}
\usepackage[numbers]{natbib}
\newtheorem{thm}{Theorem}

\title{A Factored MDP Approach to Moving Target Defense with Dynamic Threat Modeling and Cost Efficiency}

\author{
  Megha Bose\\
  \small International Institute of \small Information \\ 
  \small Technology \\
  \small Hyderabad, India \\
  \texttt{\small megha.bose@research.iiit.ac.in} \\
 \And
  Praveen Paruchuri \\
  \small International Institute of \small Information \\ 
  \small Technology \\
  \small Hyderabad, India \\
  \texttt{\small praveen.p@iiit.ac.in} \\
  \And
   Akshat Kumar\\
  \small Singapore Management University \\
  \small Singapore \\
  \texttt{\small akshatkumar@smu.edu.sg} \\
}

\begin{document}
\maketitle

\let\thefootnote\relax\footnotetext{A shorter version of this paper titled \textit{'Factored MDP based Moving Target Defense with Dynamic
Threat Modeling'} appears as an extended abstract in the \textit{Proceedings of the 23rd International Conference on Autonomous Agents and Multiagent Systems (AAMAS 2024), May 6–10, 2024, Auckland, New Zealand.}}

\begin{abstract}
Moving Target Defense (MTD) has emerged as a proactive and dynamic framework to counteract evolving cyber threats. Traditional MTD approaches often rely on assumptions about the attacker’s knowledge and behavior. However, real-world scenarios are inherently more complex, with adaptive attackers and limited prior knowledge of their payoffs and intentions. This paper introduces a novel approach to MTD using a Markov Decision Process (MDP) model that does not rely on predefined attacker payoffs. Our framework integrates the attacker’s real-time responses into the defender’s MDP using a dynamic Bayesian Network. By employing a factored MDP model, we provide a comprehensive and realistic system representation. We also incorporate incremental updates to an attack response predictor as new data emerges. This ensures an adaptive and robust defense mechanism. Additionally, we consider the costs of switching configurations in MTD, integrating them into the reward structure to balance execution and defense costs. We first highlight the challenges of the problem through a theoretical negative result on regret. However, empirical evaluations demonstrate the framework's effectiveness in scenarios marked by high uncertainty and dynamically changing attack landscapes.
\end{abstract}

\keywords{Moving Target Defense \and Markov Decision Process \and Adaptive Strategy \and Uncertainty}


\section{Introduction}

An inherent information asymmetry exists between attackers and defenders in cyber systems. This imbalance arises from the attackers' capability to conduct reconnaissance \citep{yadav2015technical} on a static system. Attackers have the freedom to systematically probe a computer network or a system and gain insights into its vulnerabilities to plan targeted attacks over a period of time. Moving Target Defense (MTD) \citep{jajodia2011moving} addresses this advantage held by attackers by introducing an increased amount of uncertainty for the attacker. This is achieved via dynamically altering the system configuration, rendering the attempts to execute successful attacks more challenging for the attacker. However, implementation of MTD solutions can introduce certain overheads like costs associated with maintaining multiple configurations and service disruptions incurred during the process of switching the system configurations. Various MTD solutions focus on altering the system with respect to what, when, and how the changes are made \citep{cho2020toward}. In the "what" aspect, solutions shuffle elements of the system, like IP addresses and software programs, or implement different versions of the system using varying technological stacks, operating systems, virtualization, etc., to create distinct system configurations. In the "when" aspect, some works explicitly address the timing of configuration changes, while others assume that the defender regularly takes switching actions on the system at predefined time intervals \citep{sengupta2017game,sengupta2020multi}. In the latter case, at the start of each time step, the defender decides whether to switch one or more adaptive aspects \citep{cho2020toward, zhuang2014towards} of the system, and concurrently, the attacker may launch an attack. The attack is executed on the configuration that results from the switching, and the attacker gains knowledge about the outcome of the attack at the end of the time step. Our work also operates under this temporal framework. The third aspect of "how" to switch the configurations has been approached via various techniques, including game theory \citep{jajodia2012moving,sengupta2017game,sengupta2020multi,zhu2013game}, reinforcement learning \citep{eghtesad2020adversarial,xu2021context,meta-rl, 8754059} and genetic algorithms \citep{liu2018hidden,makanju2017evolutionary} among others.

The MTD problem is becoming more complex in the real world as cyber systems grow in sophistication. It becomes challenging to know all the vulnerabilities in advance \citep{kamhoua2021game}. Attackers can evolve over time, and their rewards may remain unknown to the defender. In this paper, we do not assume that the attacker rewards are available as common knowledge. We model the MTD scenario as a Markov Decision Process (MDP) \citep{puterman1990markov} for the defender, considering uncertainty over the attackers. The information about exploits on the system is acquired solely through real-time interactions with the attacker, receiving information on attack success and defender losses at each time step. This interaction can be implemented using behavior fingerprinting solutions \citep{sanchez2021survey, zhang2022artificial} like User and Entity Behavior Analytics (UEBA), which can help detect and gauge the impact of a breach based on the compromised resources. We then extend the solution to a factored MDP \citep{guestrin2003efficient} for more granular modeling of the state and action spaces, enabling finer control, adaptability and more realistic representation of the aspects of the system and its dependencies. Finally, we demonstrate the effectiveness of our method across two wide domains: a web application environment using attack data from the National Vulnerability Database (NVD) and a network environment. Through empirical analysis, we showcase significant performance improvements compared to alternative methods assuming the same level of prior knowledge.

When attacker-side rewards are unknown, an adaptive attacker model learned from the real-world interactions is crucial for the defender to enhance its defense policy over time. To capture system dynamics and dependencies, we utilize a dynamic Bayesian Network (DBN) \citep{murphy2002dynamic}. The attacker's response, denoted by $\varphi$, depends on the current configuration and the switching action. In this work, we consider $\varphi$ to be a binary value that indicates the success or failure of the attack attempted in the current time step, as perceived by the defender.

\section{Related Work}

The problem of devising effective strategies for the defender in the context of MTD when facing strategic adversaries, has long garnered interest. Previous research has predominantly adopted the Stackelberg game formulation and its variations \citep{sengupta2017game, sengupta2020multi}, to model the interactions between the attacker and defender. In this model, both the players, namely the attacker(s) and defender, strive to maximize their respective rewards.

Stackelberg Security Games (SSG) \citep{dobss, kiekintveld2009computing}, have received extensive attention across a range of domains. The prevailing solution concept for these games is the Strong Stackelberg Equilibrium (SSE) \citep{breton1988sequential}, where it is assumed that the defender is aware of the defender as well as the attacker side payoffs and uses them to select an optimal mixed strategy, anticipating that the attacker will learn the mixed strategy of the defender and respond optimally. Extensions of SSG, adapted for MTD include approaches like Bayesian Stackelberg games (BSG) \citep{sengupta2017game} where a Bayesian framework is employed to capture the uncertainty regarding potential attackers. However, this approach fails to consider the multi-stage nature of the interactions. Markov game models \citep{lei2017optimal, chowdhary2019adaptive} have been proposed to address MTD scenarios, enabling state-dependent defense strategies but these models often assume a fixed attacker type. To address these limitations, \citet{sengupta2020multi} introduced the Bayesian Stackelberg Markov game (BSMG) framework. In BSMG, the attacker type adheres to a predefined distribution, and Q-learning is utilized to iteratively update value functions for both the attacker and defender and calculate the SSE at each stage, converging to the SSE of the BSMG.

In \citet{meta-rl}, a meta-reinforcement learning (meta-RL) approach has been applied to simplify the bi-level optimization problem of identifying the optimal defender solution in the presence of a strategic attacker. This simplification is achieved by introducing specific assumptions in a zero-sum Markov game, thereby reducing the complexity of finding the Strong Stackelberg Equilibrium to solving a single-agent MDP. Subsequently, meta-RL techniques have been employed to enhance the defender's policy through a training phase involving known attack scenarios, a short adaptation phase in response to real-world attacks, and a testing phase.

\citet{viswanathan2022moving} adopted a multi-armed bandit approach to address the uncertainty over attacks and adapts the Follow-the-Perturbed-Leader (FPL) algorithm with Geometric Resampling to the MTD problem. This allows for the generation of effective strategies despite having limited information about the attacker and potential exploits in the system. While \citet{policy-regret} demonstrates that no bandit algorithm can ensure sublinear regret against adaptive adversaries, empirical evidence suggests that the FPL-based method \citep{viswanathan2022moving} consistently outperforms other bandit algorithms. Furthermore, it demonstrates comparable performance to state-of-the-art methods that require prior knowledge of attacker payoffs and intentions. However, when choosing an action, the bandit model lacks the ability to account for the impact of that action on future outcomes.

Game-theoretic models are known to grapple with issues stemming from assumptions such as player rationality or the full knowledge of the payoff functions of all parties, which may not hold in many real-world MTD scenarios. Reinforcement learning approaches are frequently plagued by sample inefficiency, leading to higher exploration-exploitation costs, especially when dealing with dynamic environments. Most of the prior works are not designed for continuous adaptation and, hence, fail to adapt to evolving attacks. Table \ref{tab:mtc_comparison} in the appendix provides a concise comparison of these and related methods in various aspects. In this paper, we aim to minimize assumptions about the attacker. Emphasizing the importance of learning an attacker response model, we highlight its role in shaping the defense policy and facilitating swift adaptation to new attack scenarios. This allows the defender to derive optimal strategies based on accumulated experience while minimizing learning delays and switching costs.

\section{Problem Setting}
We view the MTD problem from the defender's perspective as an infinite horizon deterministic Markov Decision Process (MDP) $<S,A,P,R, \gamma>$. The state space $S$ of the MDP contains all possible system configurations. The action space $A$ contains the actions available to the defender corresponding to switching actions between configurations. Each action $a$ can be treated as the subsequent configuration reached after switching, as the transitions are deterministic in nature. When the defender selects a switching action at the start of a time step, the attacker within the environment chooses an attack. Our model assumes knowledge of only the defender's rewards but not of the attacker's.  Assumptions like zero-sum reward structures may not always hold as attacks that are most harmful to the system might not be in the best interests of or even be known to the attackers, who have their own intentions and abilities. In this framework, the primary source of uncertainty lies in the reward function, which is contingent on the attacker's actions. Consequently, we want to model the attacker's behavior and incorporate it into the MDP so that the defender can solve for the policy with the best-expected reward over time while considering the attacker's behavior.

\subsection{Threat Model}\label{sec:threat-model}
\begin{figure}[h]
  \centering
  \includegraphics[width=0.3\linewidth, height=0.25\linewidth]{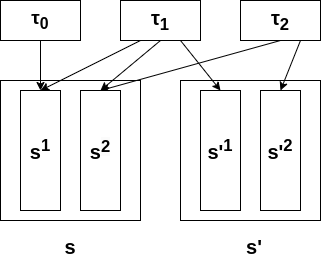}
  \caption{Representation of attacker types showing the adaptive aspects that each attacker type can attack in configurations $s$ and $s'$. For example, here, attacker type $\tau_2$ can attack adaptive aspect $s^2$ of $s$ and $s'^2$ of $s'$.\\}
  \label{fig:att-type}
\end{figure}

To effectively model the attacker, we use the concept of \textit{attacker types} \citep{meta-rl}. Let $\mathcal{T}$ denote the set of all attacker types. Each attacker type $\tau \in \mathcal{T}$ characterizes a distinct category of attackers with defined capabilities and the ability to execute a set of exploits in each configuration (see Figure \ref{fig:att-type}, Table \ref{tab:losses}) denoted by $\nu$. These attacker types may or may not be in the defender's knowledge. An attacker aligns with a unique attacker type, and all attackers within the same type exhibit similar behaviors. We consider that these attackers can demonstrate strategic behavior and may possess the knowledge of all the past configurations deployed by the defender until the current time step.

In our approach, we consider that the system confronts various attacker types, and the attacker type selection follows a fixed probability distribution for each configuration. Initially, this distribution is unknown to the defender, but the defender maintains an estimation of it using real-time interactions with the attackers. As the system encounters new attacks, the defender continuously updates its understanding of this distribution for each state. Additionally, we account for an \textit{unknown} attacker type, encompassing all the attackers not previously known to the defender. These may be stealthy attackers proficient in launching advanced or new attacks, potentially possessing specialized and superior capabilities compared to the known attacker types. Following prior work of \citet{meta-rl}, we use the concept of an attack success rate $\mu(\tau, s)$ for each attacker type $\tau$, which quantifies their proficiency in executing a successful attack in state $s$, along with an average unit time system loss $l(\tau, s)$ experienced by the defender when dealing with a successful attack from attacker type $\tau$ in state $s$. Each attack on state $s$ is characterized as $(\tau, \mu(\tau, s), l(\tau, s))$. In cases where an attacker type lacks the ability to exploit vulnerabilities within a specific configuration, their attack success probability is $0$. Consequently, when dealing with an \textit{unknown} attacker type, it is assumed to possess a high attack success rate and inflict substantial losses on the defender if it demonstrates the ability to exploit the deployed configuration.

In our model, we introduce a binary attacker response variable $\varphi \in \{0,1\}$. This variable assumes a value of $1$ if an attack proves successful and $0$ if it fails, as perceived by the defender. The outcome of this variable is contingent on the attack's success rate, which, in turn, is influenced by the attacker type that targets the system during the current time step. The attacker type is determined based on the distribution of attacker types attacking the current configuration. This mechanism governs the variability in attack success or failure within our model. In cases where the loss cannot be calculated or is unknown, the attacker type is categorized as \textit{unknown}.

\subsection{Integrating Attacker Response}
\begin{figure}[h]
  \centering
  \includegraphics[width=0.3\linewidth, height=0.25\linewidth]{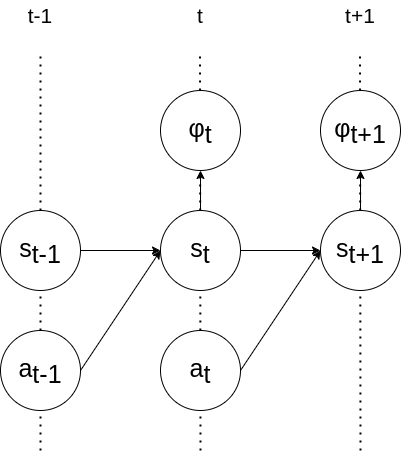}
  \vspace{1em}
  \caption{Dynamic Bayesian Network of Defender-Attacker Interactions}
  \label{fig:dbn}
\end{figure}

We employ a dynamic Bayesian Network (DBN) to incorporate the attacker response into the defender's MDP. This enables us to model the dependencies that exist among the attacker's response variable, the states, and the actions within the MDP. The success of an attack at timestep $t+1$, denoted by $\varphi_{t+1}$, depends on the configuration $s_{t+1}$ achieved through switching action $a_t$ on configuration $s_t$. It also dictates the reward $r_t$ at $t$. Figure \ref{fig:dbn} illustrates these dependencies.

\subsection{Extending to Factored MDP}
A cyber system comprises numerous attack surfaces in the adaptive aspects \citep{cho2020toward, zhuang2014towards} which contribute to the system's configuration variability and the attacker response. \citet{zhuang2014towards} presents the notion of a configuration state in MTD that captures the specific configuration of a system. This configuration state is characterized by sub-configuration parameters (or adaptive aspects of the system) that take on values from their respective domains. Examples of sub-configuration parameters can be the host memory size, hard disk size, CPU type, operating system, application software, database, programming language, IP address, open ports, etc. Building on this concept, we use a factored MDP to model these configuration parameters as state factors within the MDP. By breaking down the state and action spaces, we can enhance state representation, improve the quality of our inferences, gain finer control over switching actions, and define more complex dependencies among the adaptive aspects modeled as the state and action factors. Let $\mathcal{S}= \{ S^1, S^2, \cdots, S^n \}$ be the finite set of state variables and $\mathcal{A} = \{A^1, A^2, \cdots, A^n \}$ be the finite set of action variables. The defender's MDP is characterized by attacker-type probability estimates $\hat{P}_{att}(\tau|s, a)\, \forall \tau, s, a$ (from Algorithm \ref{algorithm:ATA-MDP-algo}) and domain information containing attack success rates $\mu(\tau, s) \, \forall \tau, s$ and unit time system losses $l(\tau, s) \, \forall \tau, s$. These are utilized to create the MDP (Algorithm \ref{algorithm:ATA-MDP-algo} step \ref{algo:create-mdp}) and set the reward values (See Eq. \ref{eq:al-phi}).

\textbf{\textit{Factored States}} - Each state factor corresponds to an adaptive aspect or sub-configuration parameter within the system. State $\boldsymbol{s}$ is represented as a tuple $(s^1, s^2, \ldots, s^n) \in Dom(\mathcal{S})$ where $s^j$ represents the value of the $j^{th}$ factor of the system's configuration $\forall j \in [n]$. For example, let the states of a system have two factors corresponding to language ($S^1$) and database ($S^2$). The domain of language factor is $Dom(S^1)=\{Python, PHP\}$ and $Dom(S^2)=\{MySQL\}$. An example state $s$ can be $(PHP, MySQL)$ where $s^1=PHP$ and $s^2=MySQL$.

\textbf{\textit{Factored Actions}} - Each action factor represents the value to which the corresponding state factor is switched to on taking the action. Similar to the state space, the action space is also factored, and each action $\boldsymbol{a}$ is represented as a tuple $(a^1, a^2, \ldots, a^n) \in Dom(\mathcal{A})$ where $a^j$ represents the value of the $j^{th}$ factor of the  new configuration that the action makes the current state switch to $\forall j \in [n]$. In the previous example, if action $a = (PHP, MySQL)$ is taken on state $s=(Python, MySQL)$, it means that the new state after switching becomes $s'=(PHP, MySQL)$.

\textbf{\textit{Factored Rewards}} - The reward consists of three parts:

1. \textit{Attack Loss ($al$)} represents the expected loss incurred by the defender from successful attacks on the system. Given probability estimate of $\hat{P}_{att}(\tau|\boldsymbol{s}, \boldsymbol{a})$ over attacker types attacking state $\boldsymbol{s}$ on taking action $\boldsymbol{a}$, the attack loss is:

\begin{equation}
\label{eq:al-phi}
al(\boldsymbol{s}, \boldsymbol{a}, \varphi) = \begin{cases}
    \frac{\sum_\tau \hat{P}_{att}(\tau|\boldsymbol{s}, \boldsymbol{a}) \mu(\tau, (\boldsymbol{s}, \boldsymbol{a})) l(\tau, (\boldsymbol{s}, \boldsymbol{a}))}{\sum_\tau \hat{P}_{att}(\tau|\boldsymbol{s}, \boldsymbol{a}) \mu(\tau, (\boldsymbol{s}, \boldsymbol{a}))} &  \varphi = 1 \\
    0 &  \varphi = 0
\end{cases}
\end{equation}

\begin{equation}
\begin{aligned}
    & al(\boldsymbol{s}, \boldsymbol{a}) = \hat{P}_{att}(\varphi=1|\boldsymbol{s}, \boldsymbol{a}) al(\boldsymbol{s}, \boldsymbol{a}, \varphi=1)\\ 
    & = \sum_\tau \hat{P}_{att}(\tau|\boldsymbol{s}, \boldsymbol{a}) \mu(\tau, (\boldsymbol{s}, \boldsymbol{a})) l(\tau, (\boldsymbol{s}, \boldsymbol{a}))
\end{aligned}
\end{equation}

Attack success probability $\mu(\tau, (\boldsymbol{s}, \boldsymbol{a})) = P(\varphi=1|\boldsymbol{s}, \boldsymbol{a}, \tau)$ and $(\boldsymbol{s}, \boldsymbol{a})$ represents the state switched to upon taking action $\boldsymbol{a}$ in state $\boldsymbol{s}$. Since transitions are deterministic, each $(\boldsymbol{s}, \boldsymbol{a})$ corresponds to a unique next state $\boldsymbol{s'}$, making $s'$ and $(s,a)$ interchangeable in this context.

2. \textit{Switching Cost ($sc$):} It accounts for the cost associated with changing the system's configuration. It is solely dependent on the system's current state $\boldsymbol{s}$ and the switching action $\boldsymbol{a}$ chosen. $\alpha$ is the relative weight given to switching costs.

3. \textit{Constant Reward ($M$):} This is a constant term reflecting the baseline reward for no attack or successful defense.

 The total reward for the defender goes as follows:

\begin{equation}
R(\boldsymbol{s}, \boldsymbol{a}) = M  - al(\boldsymbol{s}, \boldsymbol{a}) - \alpha \cdot sc(\boldsymbol{s}, \boldsymbol{a})
\end{equation}

Each of these components can be factored further into reward factors scoped on smaller subsets of $\mathcal{S} \cup \mathcal{A}$. It is to be noted that in an infinite-horizon discounted MDP, the addition of a constant reward to the reward structure does not change the optimal policy.

\section{Solution Framework}

We present our solution framework for addressing the defender's problem using an approximate linear programming (ALP) approach for solving the FMDP. Since the FMDP is unaware of the attacker type probability estimates $\hat{P}_{att}(\tau|s,a) \, \forall \tau, s, a$ (from Algorithm \ref{algorithm:ATA-MDP-algo}) and the domain information containing attack success rates $\mu(\tau, s) \, \forall \tau, s$ and unit time system losses $l(\tau, s) \, \forall \tau, s$, these are passed into the solver (Algorithm \ref{algorithm:ATA-MDP-algo} step \ref{step:algo-mdp}) to generate constraints in the primal formulation of the ALP.

\subsection{Algorithm}

To solve the defender's factored MDP augmented with the attacker's response, we use an ALP formulation for the FMDP with the constraints constructed by taking an expectation over the possible values of the binary response variable, similar to Logistic MDP \citep{mladenovlogistic} but using our probability estimates instead of the logistic regression model predictions. See Appendix \ref{appen:alp} for more details. We estimate $\operatorname{P_{att}}(\tau\mid\mathbf{s}, \mathbf{a})$ using Algorithm \ref{algorithm:ATA-MDP-algo} that calculates the estimates $\hat{P}_{att}$. 

In dynamic settings with adaptive adversaries, traditional definitions of external pseudo-regret often become inadequate. \citet{policy-regret} highlights this inadequacy and presents alternative definitions better suited for such scenarios and introduces the concept of policy regret for use in dynamic environments with adaptive adversaries. In our scenario of Moving Target Defense with switching costs, we adopt the policy regret definition and propose the following theorem. The theorem provides a negative result that guarantees that achieving sublinear regret is impossible.

\begin{thm}
\label{thm:policy-regret}
    For any MTD defense strategy on $n>1$ configurations, $\exists$ an adaptive adversary such that the defender's policy regret compared to the best static configuration in hindsight is $\Omega(T)$.
\end{thm}

\begin{figure}[h]
  \centering
  \includegraphics[width=0.6\linewidth, height=0.35\linewidth]{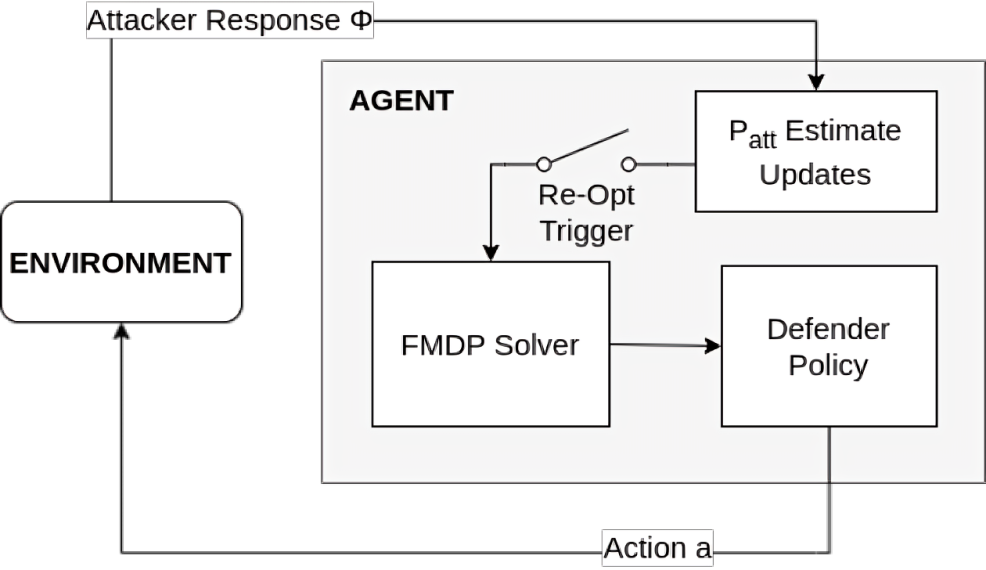}
  \vspace{1em}
  \caption{The scheme of Algorithm \ref{algorithm:ATA-MDP-algo}}
  \vspace{2em}
  \label{fig:scheme}
\end{figure}

The proof of Theorem\ref{thm:policy-regret} is delegated to Appendix \ref{appen:policy-regret}. See Appendix \ref{appen:est} and \ref{appen:regret} for analysis on the attacker type probability estimates and average regret respectively. Figure \ref{fig:scheme} shows the scheme of our solution framework.

\begin{algorithm}
  \caption{Adaptive Threat-Aware FMDP for MTD}
  \label{algorithm:ATA-MDP-algo}
  \begin{algorithmic}[1]
    \State \textbf{Initialize:} $T, \alpha, \beta, start\_state\ \boldsymbol{s_0}, M, n_{\tau, \boldsymbol{s}, \boldsymbol{a}} = 0 \, \forall \tau, \boldsymbol{s}, \boldsymbol{a}$
    \State \textbf{Load domain information} \textit{(Dom)}: states, actions, attacker types, switching costs ($sc$), attack success rates ($\mu$), unit time system losses ($l$) 
        \For{$t \gets 1$ to $T$}
         \If{re-optimization triggered}
            \State $mdp[\hat{P}_{att}, Dom] \gets create\_mdp(\hat{P}_{att}, Dom)$ \label{algo:create-mdp}
            \State $\pi \gets lp\_solver(mdp[\hat{P}_{att}, Dom], P_{att}, Dom)$ \Comment{Re-calculate FMDP policy $\pi$}\label{step:algo-mdp}
        \EndIf
          \State $\boldsymbol{a_t} \gets \pi$ \Comment{Using last calculated policy}\label{step:algo-mdp-act}
          \State $\varphi_{t+1}, \tau_t, \boldsymbol{s_{t+1}} \gets mtd\_sim(\boldsymbol{s_t}, \boldsymbol{a_t})$ \Comment{Real-world Interaction}\label{step:algo-mtd-sim}
          \State As $\boldsymbol{s_{t+1}}$ is fixed given $(\boldsymbol{s_t}, \boldsymbol{a_t})$, we use them interchangeably.
          \State $r_t \gets M - \mathbb{1}[\varphi_{t+1} = 1] l(\tau_t, (\boldsymbol{s_t}, \boldsymbol{a_t})) - \alpha \cdot sc(\boldsymbol{s_t}, \boldsymbol{a_t})$ \label{step:algo-reward}
          \State $n_{\tau, \boldsymbol{s}, \boldsymbol{a}} \gets \frac{n_{\tau, \boldsymbol{s}, \boldsymbol{a}}}{\beta} + \mathbb{1}[\varphi_{t+1} = 1, \tau = \tau_t, \boldsymbol{s} = \boldsymbol{s_t}, \boldsymbol{a} = \boldsymbol{a_t}] \, \forall \tau, \boldsymbol{s}, \boldsymbol{a}$
          \Comment{Eq. \ref{eq:n}}\label{step:algo-num-success}
          \State $\hat{P}_{att}(\tau | \boldsymbol{s}, \boldsymbol{a}) \gets \begin{cases} 
              \frac{1}{\mathcal{N}} \frac{n_{\tau, \boldsymbol{s}, \boldsymbol{a}}}{\mu(\tau, (\boldsymbol{s}, \boldsymbol{a}))} & \text{if } \mu(\tau, (\boldsymbol{s}, \boldsymbol{a})) > 0 \\
              0 & \text{otherwise}
           \end{cases}\, \forall \tau, \boldsymbol{s}, \boldsymbol{a}$ \Comment{Eq. \ref{eq:p-att}} \label{step:algo-p-att}
        \EndFor

  \end{algorithmic}
\end{algorithm}

The algorithm starts with the initialization of hyperparameters and variables and the loading of domain information. Then, we run the algorithm for $T$ timesteps. In each timestep, if anomalous activity is detected, we run the re-optimization protocol in Step \ref{step:algo-mdp} to update the FMDP policy based on the current estimate of attacker type probabilities. The defender takes an action according to the last calculated policy (Step \ref{step:algo-mdp-act}), and based on that, it receives the attacker response value $\varphi$, the attacker type ($unknown$ type if the attacker type is unknown to the defender), and the next state (Step \ref{step:algo-mtd-sim}). Using the attack response and unit-time system loss $l(\tau, (\boldsymbol{s}, \boldsymbol{a}))$ due to the attacker type $\tau$ (tuple $(\boldsymbol{s},\boldsymbol{a})$ and next state $\boldsymbol{s'}$ have the same meaning as transitions are deterministic) and the switching cost incurred in executing the current switching action, we calculate the defender's reward. We maintain a temporally weighted value $n$ for each attacker type, state, and action tuple, representing a weighted sum of the attack type's success in that state when that action was executed using a weighing factor $\beta$. At timestep $t$, $n$ is updated $\forall \tau, \boldsymbol{s}, \boldsymbol{a}$  as follows in Step \ref{step:algo-num-success}. $\mathbb{1}$ is the indicator function. If $\beta > 1$, the weightage of earlier successes reduces with every timestep. This is used to update the attacker-type probability estimate in Step \ref{step:algo-p-att}:

\begin{equation}
    \begin{aligned}
    \label{eq:n}
    n_{\tau, \boldsymbol{s}, \boldsymbol{a}} \gets \frac{n_{\tau, \boldsymbol{s}, \boldsymbol{a}}}{\beta}+ \mathbb{1}[\varphi_{t+1} = 1, \tau = \tau_t, \boldsymbol{s} = \boldsymbol{s_t}, \boldsymbol{a} = \boldsymbol{a_t}]
    \end{aligned}
\end{equation}

\begin{equation}
    P_{att}(\tau|\boldsymbol{s}, \boldsymbol{a}) = \frac{P(\tau, \varphi = 1, \boldsymbol{s}, \boldsymbol{a})}{P(\varphi=1|\boldsymbol{s}, \boldsymbol{a}, \tau) P(\boldsymbol{s}, \boldsymbol{a})}
\end{equation}

Hence, given state $s$ and action $a$, we have

\begin{equation}
\label{eq:p-att}
P_{att}(\tau|\boldsymbol{s}, \boldsymbol{a}) \propto \frac{n_{\tau, \boldsymbol{s}, \boldsymbol{a}}}{\mu(\tau, (\boldsymbol{s}, \boldsymbol{a}))} \implies
\hat{P}_{att}(\tau|\boldsymbol{s}, \boldsymbol{a})
= \frac{1}{\mathcal{N}}\frac{n_{\tau, \boldsymbol{s}, \boldsymbol{a}}}{\mu(\tau, (\boldsymbol{s}, \boldsymbol{a}))}
\end{equation}

where $n_{\tau, \boldsymbol{s}, \boldsymbol{a}}$ is a temporally weighted estimate of the number of attack successes that occurred in the state $s$, on action $a$ and from attacker type $\tau$ while attack success rate $\mu(\tau, (\boldsymbol{s}, \boldsymbol{a}))$ represents the inherent proficiency of the attacker type in making a successful attack attempt given the state and action and $\mathcal{N}$ is the normalizing factor. We assign a $\mu$ of $1$ to the $unknown$ attacker type since the defender lacks information about its true proficiency.

\section{Empirical Evaluation}

\subsection{Environmental Setup}

We look at a web application and a network-based environment to showcase the performance of MTD algorithms. (See Appendix \ref{appen:domain-diag} for diagrams)

\textbf{Web Application Environment} Inspired by previous works \citep{sengupta2017game, meta-rl, li2020spatial}, we employ the National Vulnerability Database (NVD) data from the years 2020 to 2022 and Common Vulnerability Scoring System (CVSS) scores to establish the experimental framework for a web application. We define the set of system configurations as \{($PHP$, $MySQL$), ($Python$, $MySQL$), ($PHP$, $PostgreSQL$), ($Python$, $PostgreSQL$)\}. Here, the first factor specifies the programming language, and the second factor represents the database used in the application. Hence, we have a domain with $2$ state factors each with a domain size of $2$ and $2$ action factors each with a domain size of $2$, which leads to $8$ sparse binarized features (sbfs) and induced state and action spaces of size $4$ each. Actions correspond to configurations to which the current state will transition to, as defined earlier.

Rewards are computed using the Common Vulnerability Scoring System (CVSS) scores. Each attacker type can exploit a specific set of vulnerabilities. Following \citet{meta-rl}, for a given state, the attack success rate ($\mu$) for a particular attacker type in a given state is determined by the average exploitability score (ES) of the vulnerabilities that the attacker type can exploit in that state, calculated by $0.1 * ES$ where $ES \in [0, 10]$. The unit time system loss ($l$) is calculated based on the average impact score (IS) of the vulnerabilities that the attacker type can exploit in the configuration, calculated by $-10 * IS$ where $IS \in [0, 10]$. Based on the probability distribution over attacker types that can attack the current state, one of the attacker types attempts an attack. The outcome of the attack depends on the configuration switched to and the attacker's known success rate for that attacker type. Here, three attacker types are considered: Mainstream Hacker ($MH$), Database Hacker ($DH$) and $unknown$. The switching costs ($sc$) used are given in Table \ref{tab:sc} and the attacker capabilities, attack success rates ($\mu$) and unit time system losses ($l$) are given in Table \ref{tab:losses}.

\begin{figure}[h]
\begin{small}
  \begin{subfigure}{0.35\textwidth}
    \caption{Switching Costs}
    \vspace{1em}
        \centering
        \label{tab:sc}
        \begin{tabular}{ccccc}\toprule
            \textit{} & $C_1$ & $C_2$ & $C_3$ & $C_4$ \\ \midrule
            $C_1$ & 0 & 20 & 60 & 100 \\
            $C_2$ & 20 & 0 & 90 & 50 \\
            $C_3$ & 60 & 90 & 0 & 20 \\
            $C_4$ & 100 & 50 & 20 & 0 \\\bottomrule
        \end{tabular}
    \vspace{1em}
  \end{subfigure}
\end{small}
  \begin{small}
  \begin{subfigure}{0.65\textwidth}
    \caption{Attacker Type Capabilities, Attack Success Rates, and Unit Time System Losses (true values for the \textit{unknown} (unk) type)\\}
	\label{tab:losses}
	\begin{tabular}{lllll}\toprule
		\textit{} & $C_1$ & $C_2$ & $C_3$ & $C_4$ \\ \midrule
	    $\nu_{MH}$ & \{PHP, MySQL\} & \{Python, MySQL\} & \{PHP\} & \{Python\}\\	  
            $\mu_{MH}$ & 0.32 & 0.32 & 0.36 & 0.36 \\
		  $l_{MH}$ & 61 & 43 & 66 & 29 \\ 
            \midrule
            $\nu_{DH}$ & \{MySQL\} & \{MySQL\} & \{Postgres\} & \{Postgres\}\\
		$\mu_{DH}$ & 0.7 & 0.7 & 0.65 & 0.65 \\
		  $l_{DH}$ & 43 & 43 & 50 & 50 \\
            \midrule
            $\nu_{unk}$ & \{PHP, MySQL\} & \{MySQL\} & \{PHP\} & \{\}\\	  
            $\mu_{unk}$ & 0.78 & 0.7 & 0.87 & 0.0 \\
		  $l_{unk}$ & 100 & 100 & 100 & 0 \\ 
            \bottomrule
	\end{tabular}
  \end{subfigure}
\end{small}
\vspace{1em}
\caption{In the web application environment, configurations $C_1 = (PHP, MySQL)$, $C_2 = (Python, MySQL)$, $C_3 = (PHP, Postgres)$, $C_4 = (Python, Postgres)$\\}
\vspace{1em}
\end{figure}

\textbf{Network Environment} For simulating and testing network security in general, we have tools like Microsoft's CyberBattleSim \citep{msft:cyberbattlesim}. However, applying MTD on all nodes is often costly and sometimes not even possible. Hence, MTD can be applied on critical nodes like servers \citep{eghtesad2020adversarial}. Here, we focus on a network with two chosen nodes where MTD is deployed. The state representation employs a binary vector of length $2$ (two nodes), with each value denoting the online ($1$) or offline ($0$) status of each node. Actions determine the subsequent state transition, as defined earlier. The defender suffers from a switching cost of $50$ for each node transitioning to offline mode. Inspired by CyberBattleSim attacks, we classify attacker types based on $(source\,node, target\, node)$ tuples. A distinction is drawn between local attack types, where the source is equal to the target, and remote attack types, where the source and target differ, and finally, an unknown attack type, leading to five attack types. Attack success rates ($\mu$) for local attacks are chosen from $\mathcal{U}(0.5, 0.6)$, while remote attacks exhibit a success rate chosen from $\mathcal{U}(0.2, 0.3)$, aligning with the inherent challenges of launching successful attacks remotely. Unit time system losses ($l$) chosen from $\mathcal{U}(60, 70)$ are incurred on successful attacks. It is to be noted that the number of constraints in the ALP is not dependent on the number of attacker types considered. Hence, according to the requirements and specific network properties, the attacker-type definition can be expanded to include vulnerability sets in the tuple besides the source and target node.

In both environments, we do not use any special event-triggered adaptation. Instead, at every timestep, the attacker-type probability distribution is updated using Bayesian inference and the experience gained in all the previous timesteps is incorporated into the defender's MDP. This approach keeps the model up-to-date with evolving attack patterns while maintaining stability.

\subsection{Defense Methods Compared} 
Our approach, which we call 'Adaptive Threat-Aware Factored MDP' ($ATA-FMDP$), is compared to other methods that utilize a similar amount of prior information. Specifically, we compare it against the following three approaches: (a) A bandit-based approach ($FPL-MTD$) \citep{viswanathan2022moving} - $FPL-MTD$ has proved as a strong baseline when compared with other approaches that do not consider prior knowledge regarding attackers. (b) $EPS-GREEDY$ approach - this approach uses epsilon-greedy based exploration-exploitation where for exploitation it greedily selects the action that has the highest action-value estimate, and (c) a uniform random defense ($URS$) policy that selects next actions in a uniformly random manner. We conduct our experiments in $10$ iterations, each with $1000$ timesteps. $\beta$ is $2$, $M$ is $200$ and discount factor is $0.9$. Epsilon of $EPS-GREEDY$ is $0.2$. The hyperparameters of $FPL-MTD$ are taken from \citet{viswanathan2022moving}

\subsection{Response to Evolving Attack Landscapes}
\label{sec:5-2}
In the real-world dynamic attack landscape, attackers acquire knowledge gradually, learning when their attacks are most likely to yield rewards. They may also observe the defender's actions over time and decide when to launch attacks on the system. In this case, the distribution of attacker types within each system state is not uniform across all states; it evolves continuously. For instance, a state with robust defenses capable of thwarting most attacks will not attract attacks from rational attackers. Conversely, say in the web application environment, if a state has vulnerabilities in the database, the probability of a database hacker launching an attack on that state is expected to be higher.

\begin{figure*}[h]
  \centering
  \begin{subfigure}{0.3\linewidth}
    \centering
    \includegraphics[width=\linewidth]{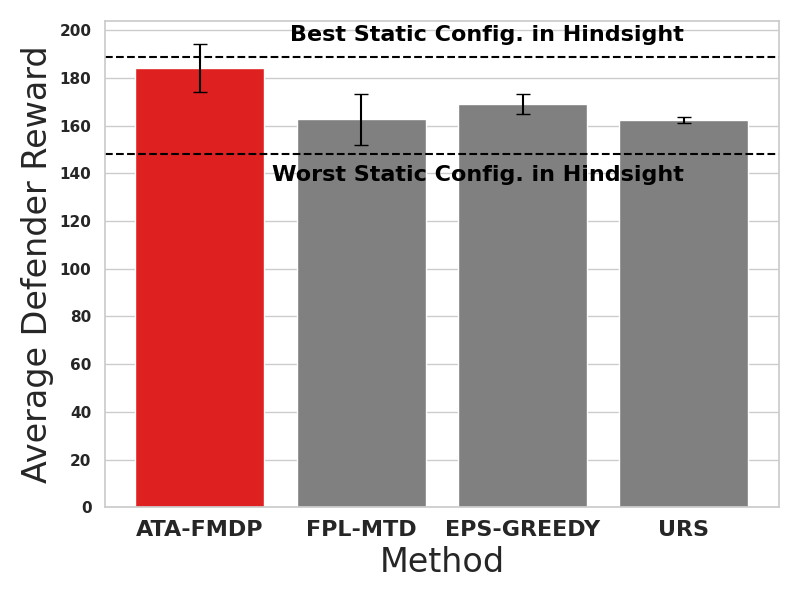}
    \caption{$\alpha = 0$}
    \vspace{1em}
    \label{fig:evolve-stota}
  \end{subfigure}
  \hfill
  \begin{subfigure}{0.3\linewidth}
    \centering
    \includegraphics[width=\linewidth]{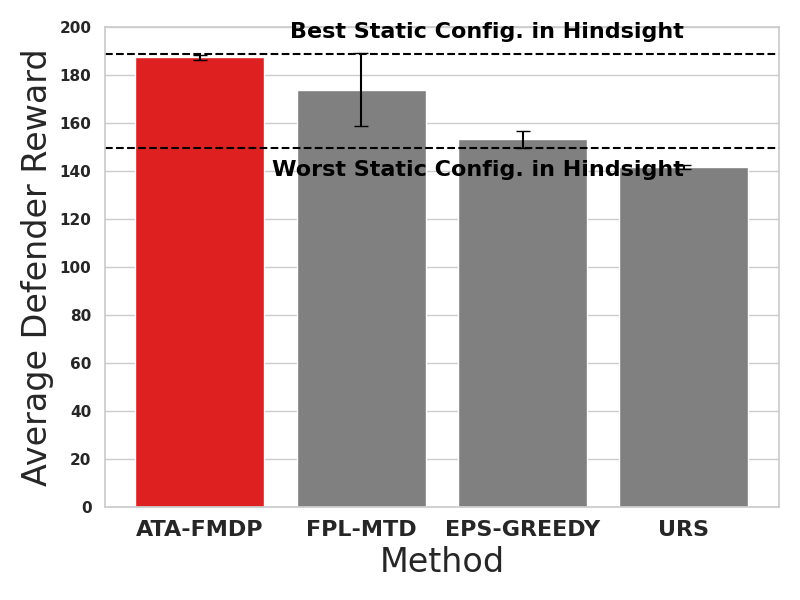}
    \caption{$\alpha = 0.5$}
    \vspace{1em}
    \label{fig:evolve-stotb}
  \end{subfigure}
  \hfill
  \begin{subfigure}{0.3\linewidth}
    \centering
    \includegraphics[width=\linewidth]{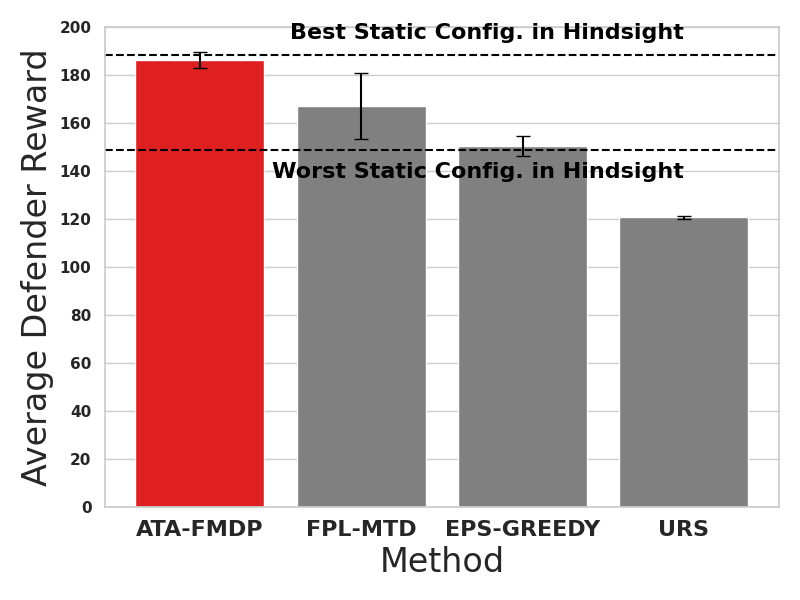}
    \caption{$\alpha = 1.0$}
    \vspace{1em}
    \label{fig:evolve-stotc}
  \end{subfigure}
  
  \begin{subfigure}{0.3\linewidth}
    \centering
    \includegraphics[width=\linewidth, height=0.7\linewidth]{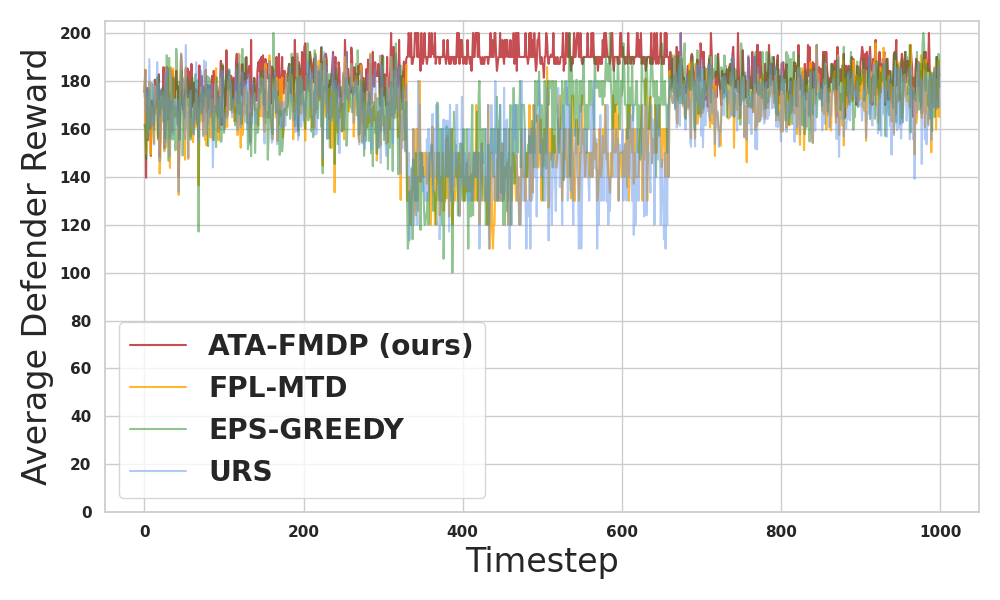}
    \caption{$\alpha = 0$}
    \vspace{1em}
    \label{fig:evolve-stotd}
  \end{subfigure}
  \hfill
  \begin{subfigure}{0.3\linewidth}
    \centering
    \includegraphics[width=\linewidth, height=0.7\linewidth]{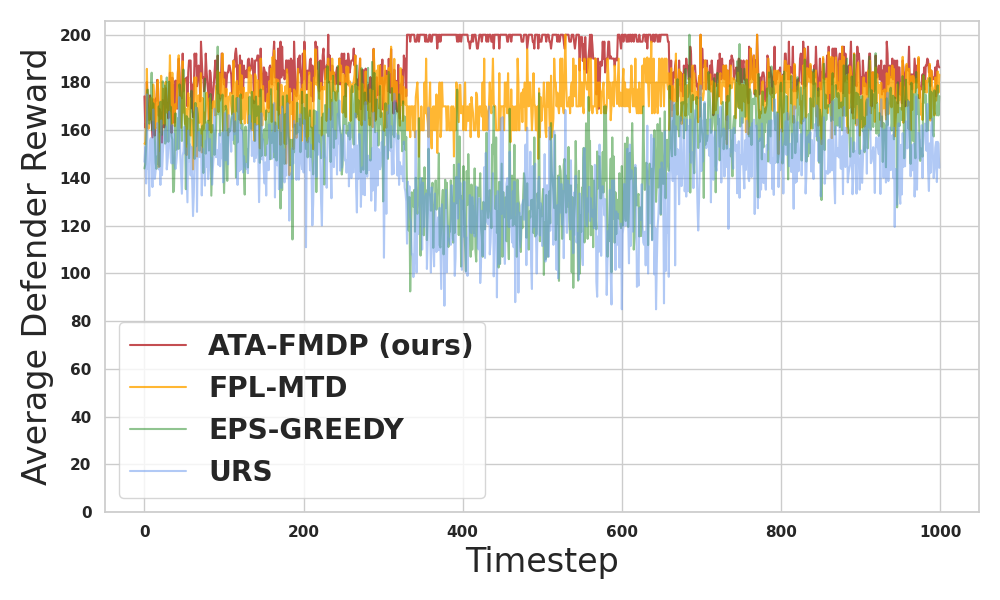}
    \caption{$\alpha = 0.5$}
    \vspace{1em}
    \label{fig:evolve-stote}
  \end{subfigure}
  \hfill
  \begin{subfigure}{0.3\linewidth}
    \centering
    \includegraphics[width=\linewidth, height=0.7\linewidth]{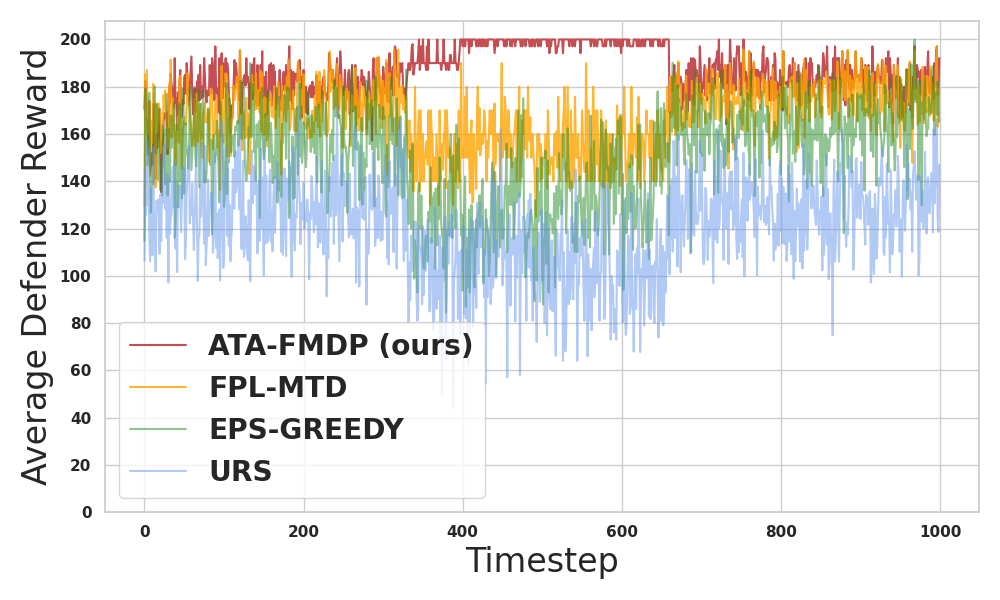}
    \caption{$\alpha = 1.0$}
    \vspace{1em}
    \label{fig:evolve-stotf}
  \end{subfigure}
  \vspace{1em}
  \caption{Web Application Environment - Evolving Attack Landscape where between timesteps $330$ and $660$, the $unknown$ attacker prevails. Graphs (a,b,c) show cumulative defender rewards for each defender strategy, while (d,e,f) show the evolution of defender rewards over the $1000$ timesteps. $\alpha$ refers to the relative weight given to switching costs.}
  \vspace{1em}
  \label{fig:evolve-stot}
\end{figure*}

\begin{figure}[h]
  \centering
  \begin{subfigure}{0.4\linewidth}
    \centering
    \includegraphics[width=0.9\linewidth]{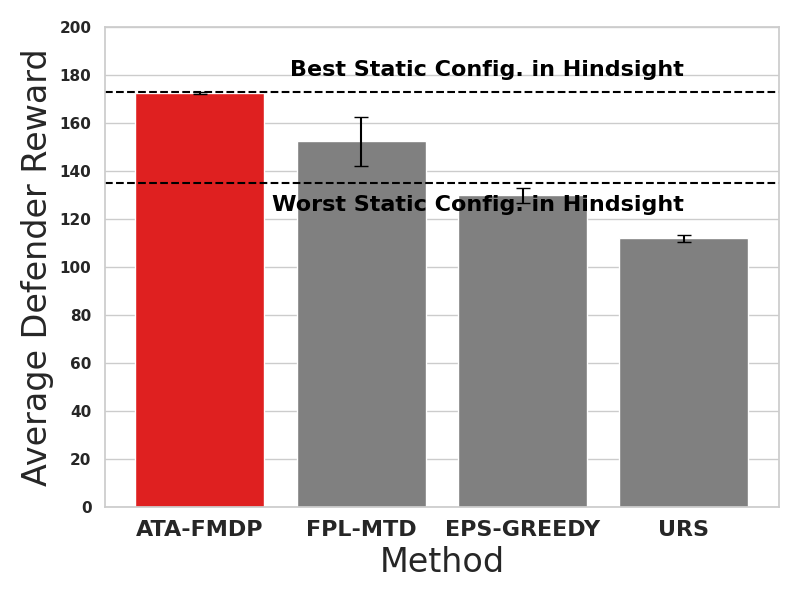}
    \caption{}
    \vspace{1em}
    \label{fig:most-adverse-evolvec}
  \end{subfigure}
  \begin{subfigure}{0.4\linewidth}
    \centering
    \includegraphics[width=0.85\linewidth, height=0.65\linewidth]{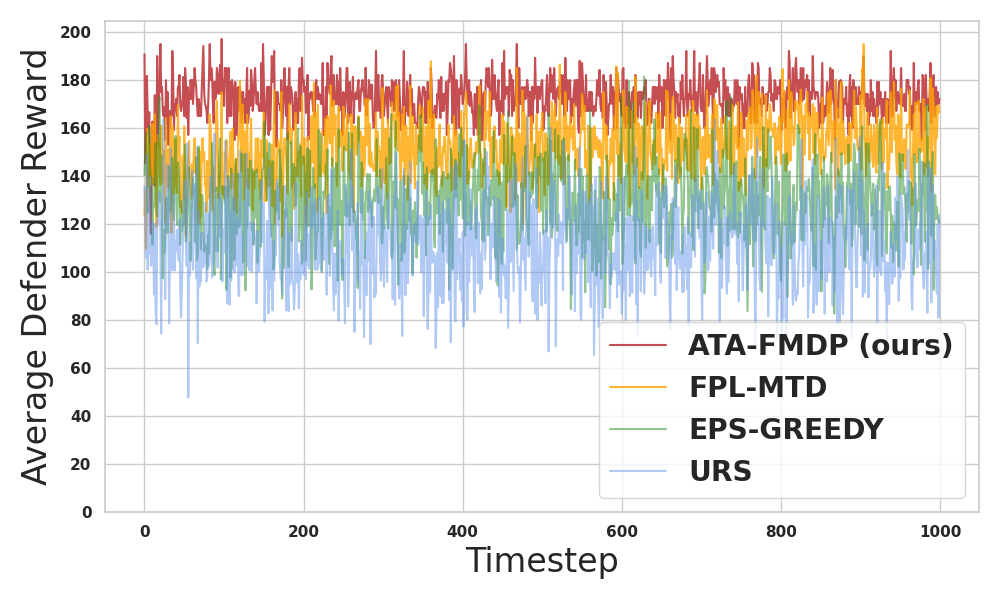}
    \caption{}
    \vspace{1em}
    \label{fig:most-adverse-evolvef}
  \end{subfigure}
  \vspace{1em}
  \caption{Web Application Environment - Evolving Attack Landscape with strategic attackers that resort to the most adverse strategy based on defender policy estimated till current timestep. Graph (a) shows cumulative defender rewards for each defender strategy, while (b) shows the evolution of defender rewards over the $1000$ timesteps. Here, $\alpha=1$.\\}
  \vspace{1em}
  \label{fig:most-adverse-evolve}
\end{figure}

We compare the average defender reward received using each of the defense strategies. We analyze the rewards averaged over 10 iterations, each consisting of 1000 timesteps. In the figures, the y-axis represents this average reward, while the x-axis represents the defense strategy. The dotted lines indicate the average defender rewards that could have been achieved if the best and worst static defenses in hindsight were deployed. These lines provide a benchmark for assessing the effectiveness of the switching strategy against the best possible performance a static defense could achieve, had it known the optimal configuration in advance, and against the worst-case performance if the least effective configurations were deployed without any switching. We also present the evolution of the defender rewards averaged over the $10$ iterations through $1000$ timesteps. The y-axis represents this average reward and the x-axis represents the timesteps.

\textbf{In Web Application Environment } We consider an evolving attack landscape defined by initial probabilities of $\{0.5, 0.35, 0.15\}$ for $MH$, $DH$, and $unknown$ attackers, which change to $\{0.1, 0.0, 0.9\}$ for $MH$, $DH$, and $unknown$ attacker types between timesteps $330$ and $660$ and then back to initial probabilities. This kind of situation can occur if an $unknown$ attacker enters the landscape with targeted attacks. The $unknown$ attacker type can cause significant harm if the defender's strategy doesn't adapt to the landscape in time. From figures \ref{fig:evolve-stota}, \ref{fig:evolve-stotb}, \ref{fig:evolve-stotc}, we see that on average, our $ATA-FMDP$ approach performs $~13\%$, $~34\%$ and $~56\%$ better than $FPL-MTD$, $EPS-GREEDY$ and $URS$, respectively. From Figures \ref{fig:evolve-stotd}, \ref{fig:evolve-stote}, \ref{fig:evolve-stotf}, we observe that our approach quickly adapts to the new scenario and maintains a high defender reward, while the other methods experience a drastic drop in the defender reward when faced with the strong unknown attacker. This demonstrates the robustness of our approach in evolving attack landscapes.

The attack landscape might not just be evolving but can also contain strategic attackers that observe the defender's actions over time, and keep changing their strategy based on observations. We consider such an evolving attack landscape with strategic attackers where the attackers execute actions that cause the most harm to the defender at each timestep as the most adverse scenario for the defender. This is different from the best response attack strategy since we do not assume knowledge about the attacker's side intentions and do not consider the scenario to be zero-sum. In the adverse attack scenario, the attackers use the estimated defender policy they have observed until now, and then based on the current configuration and the estimated defender policy, they deploy the attack most adverse to defender. From Figure \ref{fig:most-adverse-evolvec}, we derive that on average, our $ATA-FMDP$ approach performs $~15\%$, $~20\%$ and $~52\%$ better than $FPL-MTD$, $EPS-GREEDY$ and $URS$, respectively. From Figure \ref{fig:most-adverse-evolvef}, we see that it gives better rewards than other methods across the $1000$ timesteps.

\begin{figure}[h]
  \centering
  \begin{subfigure}{0.47\linewidth}
    \centering
    \includegraphics[width=0.75\linewidth]{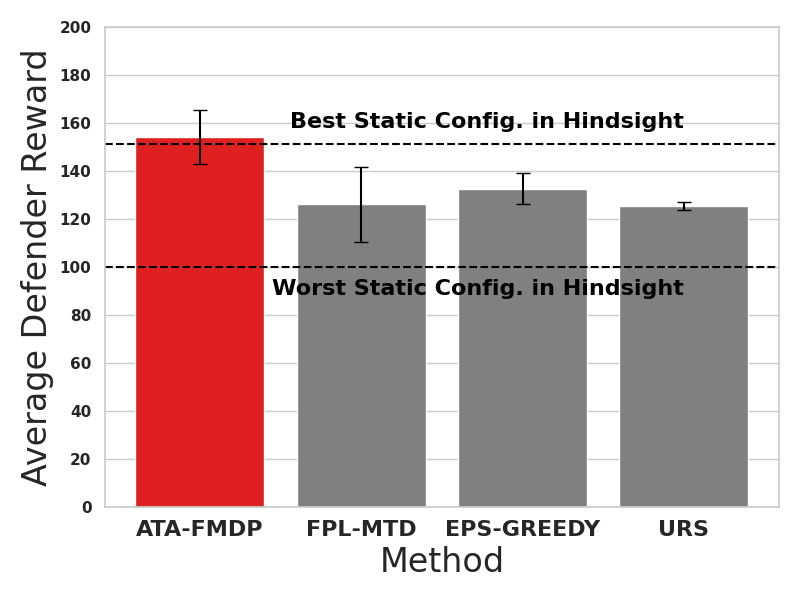}
    \caption{}
    \vspace{1em}
    \label{fig:net-evolve-stotc}
  \end{subfigure}
  \begin{subfigure}{0.48\linewidth}
    \centering
    \includegraphics[width=0.75\linewidth, height=0.5\linewidth]{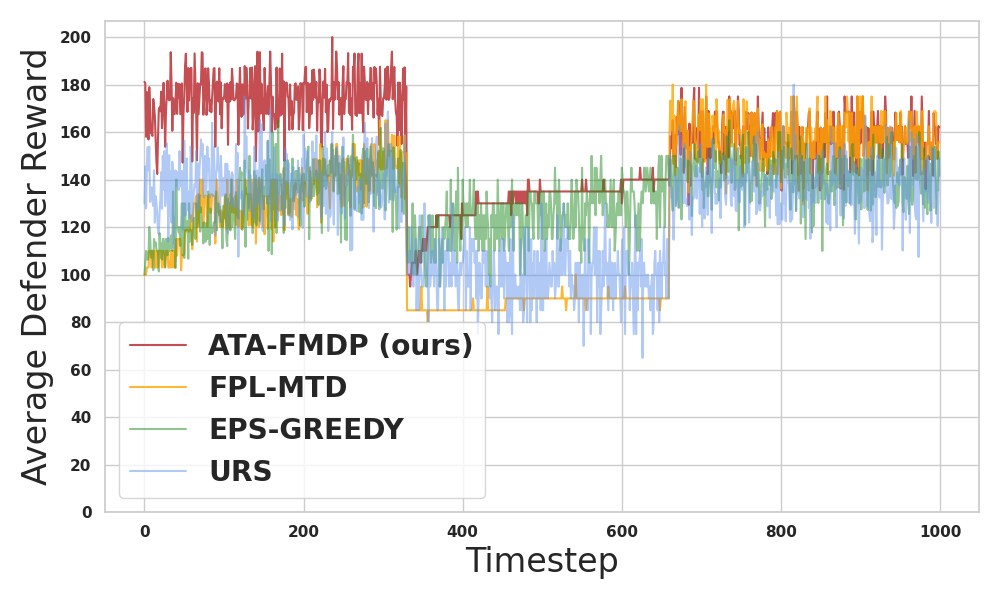}
    \caption{}
    \vspace{1em}
    \label{fig:net-evolve-stotf}
  \end{subfigure}
  \begin{subfigure}{0.47\linewidth}
    \centering
    \includegraphics[width=0.75\linewidth]{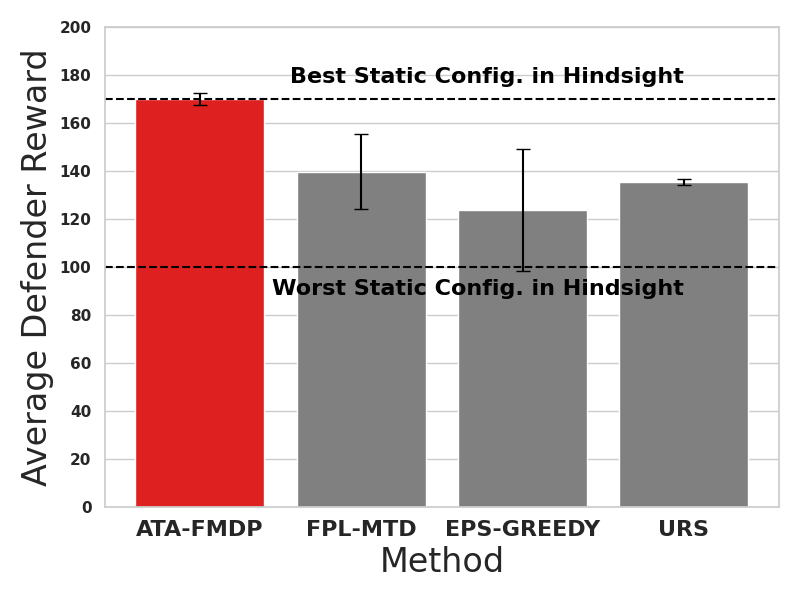}
    \caption{}
    \vspace{1em}
    \label{fig:net-most-adverse-evolvec}
  \end{subfigure}
  \begin{subfigure}{0.5\linewidth}
    \centering
    \includegraphics[width=0.75\linewidth, height=0.5\linewidth]{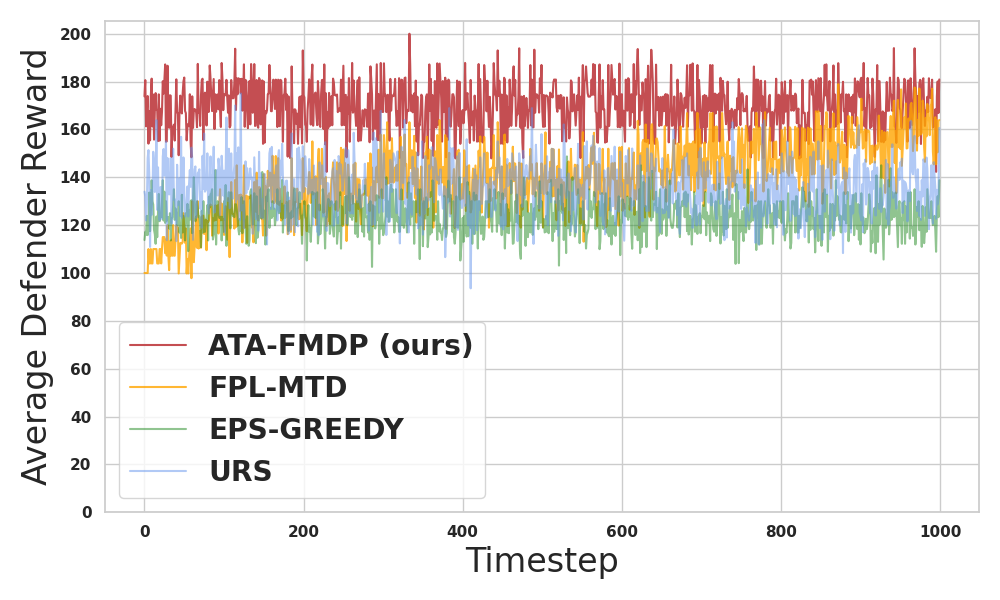}
    \caption{}
    \vspace{1em}
    \label{fig:net-most-adverse-evolvef}
  \end{subfigure}
  \vspace{1em}
  \caption{Network Environment - (a, b) - Evolving Attack Landscape where between timesteps $330$ and $660$, the $unknown$ attacker prevails. (c, d) - Evolving Attack Landscape with strategic attackers using the most adverse strategy based on defender policy estimated till current timestep.\\}
  \label{fig:most-adverse-evolve-network}
  \vspace{1em}
\end{figure}

\textbf{In Network Environment } We first consider an attack lanscape that is characterised by attacker type probabilities of $\{0.2, 0.3, 0.3, 0.2\}$ over the attacker types $\{(0, 0), (0, 1), (1, 0), (1, 1)\}$ wherein attacker type $(a, b)$, $a$ is the source node and $b$ is the target node. Then, between timesteps $330$ and $660$, an unknown attacker that is capable of causing attack damage of $100$ on Node $0$ prevails. In this phase, it is easy to see that the optimal approach would be to keep Node $0$ offline to avoid major damage. As we see from Figure \ref{fig:net-evolve-stotc}, our approach $ATA-FMDP$ performs $18\%$ better than $FPL-MTD$, $15\%$ better than $EPS-GREEDY$ and $23\%$ better than $URS$ in this scenario. From Figure \ref{fig:net-evolve-stotf}, we can see how the average reward evolves and in the phase where unknown attacker prevails, it is seen that $ATA-FMDP$ keeps Node $0$ offline as much as possible, which is also reflected by the unknown attacker probability estimates and the average reward in those timesteps. This is not seen in other methods.

Similar to the most adverse attack scenario of web application, we also implement a case in the network where attacker avails strategy most adverse to the defender and we see from Figure \ref{fig:net-most-adverse-evolvec} that our approach $ATA-FMDP$ performs $26\%$ better than $FPL-MTD$, $12\%$ better than $EPS-GREEDY$ and $26\%$ better than $URS$.

In Figure \ref{fig:net-evolve-stotf}, we see that $FPL-MTD$ struggles to adapt to the unknown attacker quickly. In Figure \ref{fig:net-most-adverse-evolvef}, we see that $FPL-MTD$ shows slower convergence than $EPS-GREEDY$ under the most-adverse attack scenario in the network domain and hence gives lower reward on average. However, it eventually catches up. It is seen that despite lacking sophisticated adaptability, $URS$ remains effective in some dynamic attack environments characterized by strategic adversaries. Intelligent approaches may struggle to adapt to evolving conditions swiftly, adhering to outdated patterns, and potentially offering attackers opportunities for exploitation. $ATA-FMDP$ proves to be effective in both the cases. It manages to take actions to keep Node 0 offline by understanding the unknown attacker's attack pattern in the first case and it is also able to maintain high average reward in the most adverse attack landscape.

\subsection{Impact of Switching Cost}
It is known that in certain systems, switching costs tend to be generally lower, while in others, the process of making switches can be considerably challenging. We vary the value of the switching cost weighing factor $\alpha$ among $\{0.0, 0.5, 1.0\}$ in our analysis and show the results on the web application environment in Figure \ref{fig:evolve-stot}. We observe that with an increased weightage ($\alpha$) given to switching costs, $URS$ performs much worse than other methods. This is expected because $URS$ is a cost-agnostic method and hence suffers from more switching costs. We also notice that with increasing $\alpha$, the methods other than our approach perform worse. This occurs due to the short-term optimization of these methods i.e., they aim to maximize immediate outcomes and hence are reluctant to switch due to the higher switching costs. However, in the long term, switching to the state $\{Python, PostgreSQL\}$ proves beneficial during the period of the prevailing $unknown$ attacker from timestep $330$ to $660$, since it is the safest state given the attack landscape. Reluctance to switching due to a lack of foresight leads to reduced adaptability, and hence, our approach outperforms the other methods.

\subsection{Adaptability and Other Insights}
The model's significant feature lies in its capacity to effectively handle unknown attacks. In cases where a system encounters an unfamiliar and potentially stealthy attack characterized by a potentially high attack rate and impact, the model is designed to detect and respond to it promptly. The defender's policy adapts to mitigate the threat until more information becomes available about the nature of the attack and its intentions. Additionally, an intriguing analysis pertains to the prevalence of specific actions chosen within the system by the defender. This exploration provides insights into the underlying dynamics of the system configurations and the broader attack landscape. Prolonged avoidance of a specific configuration may signal an underlying issue or an inherent vulnerability within that configuration. We also analyze an unknown attack landscape involving a specialized database hacker targeting only PostgreSQL in Appendix \ref{appen:dh-unknown}. A detailed analysis of switching patterns in the action factors indicated potential vulnerability in PostgreSQL as expected.

\section{Conclusions and Future Work}
In this work, we modeled two broad real-world Moving Target Defense (MTD) scenarios with minimal assumptions about attackers and no prior knowledge of the attacker's rewards and intentions. Our approach introduces an attacker model that learns from real-time attacks and continuously updates attack success probabilities, seamlessly integrating this information into the defender's Markov Decision Process (MDP). The defender determines an optimal policy by solving the MDP formulation using the latest attacker response predictions. These estimates undergo periodic updates to maintain the model's accuracy. Our method compares favorably to the other approaches, assuming a similar level of information regarding the evolving and adaptive attacker settings in two different domains. Notably, our approach demonstrates the capability to model unknown attacker types and adapt to new and unfamiliar attack patterns. Insights into the system and attack landscape can be extracted from the calculated policies. We also look at the sensitivity of our approach to different importance weights assigned to switching costs and draw pertinent conclusions. Even with limited initial knowledge of the attack landscape, the defender acquires intelligent insights and adapts favorably to reduce losses from attacks and avoidable costly configuration changes.

\textbf{Future Work } Following directions can be pursued based on this work: Incorporation of more advanced attack success predictors, such as Artificial Neural Networks designed to handle sequential data, can enhance the learning of attacker behaviors over time. Within the realm of unknown attackers, the application of clustering methods based on the losses incurred due to their attacks can help incrementally refine the threat model by grouping similar attacker types. Testing the proposed method on real-world systems is another potential direction. In such scenarios, rather than relying solely on standard scoring systems or synthetic data, system-specific rewards and factors such as blast radius can be incorporated to assess the severity of the breaches more accurately.


\bibliographystyle{unsrtnat}  
\bibliography{references}

\begin{thebibliography}{34}
\providecommand{\natexlab}[1]{#1}
\providecommand{\url}[1]{\texttt{#1}}
\expandafter\ifx\csname urlstyle\endcsname\relax
  \providecommand{\doi}[1]{doi: #1}\else
  \providecommand{\doi}{doi: \begingroup \urlstyle{rm}\Url}\fi

\bibitem[Yadav and Rao(2015)]{yadav2015technical}
Tarun Yadav and Arvind~Mallari Rao.
\newblock Technical aspects of cyber kill chain.
\newblock In \emph{Security in Computing and Communications: Third International Symposium, SSCC 2015, Kochi, India, August 10-13, 2015. Proceedings 3}, pages 438--452. Springer, 2015.

\bibitem[Jajodia et~al.(2011)Jajodia, Ghosh, Swarup, Wang, and Wang]{jajodia2011moving}
Sushil Jajodia, Anup~K Ghosh, Vipin Swarup, Cliff Wang, and X~Sean Wang.
\newblock \emph{Moving target defense: creating asymmetric uncertainty for cyber threats}, volume~54.
\newblock Springer Science \& Business Media, 2011.

\bibitem[Cho et~al.(2020)Cho, Sharma, Alavizadeh, Yoon, Ben-Asher, Moore, Kim, Lim, and Nelson]{cho2020toward}
Jin-Hee Cho, Dilli~P Sharma, Hooman Alavizadeh, Seunghyun Yoon, Noam Ben-Asher, Terrence~J Moore, Dong~Seong Kim, Hyuk Lim, and Frederica~F Nelson.
\newblock Toward proactive, adaptive defense: A survey on moving target defense.
\newblock \emph{IEEE Communications Surveys \& Tutorials}, 22\penalty0 (1):\penalty0 709--745, 2020.

\bibitem[Sengupta et~al.(2017)Sengupta, Vadlamudi, Kambhampati, Doup{\'e}, Zhao, Taguinod, and Ahn]{sengupta2017game}
Sailik Sengupta, Satya~Gautam Vadlamudi, Subbarao Kambhampati, Adam Doup{\'e}, Ziming Zhao, Marthony Taguinod, and Gail-Joon Ahn.
\newblock A game theoretic approach to strategy generation for moving target defense in web applications.
\newblock In \emph{AAMAS}, volume~1, pages 178--186, 2017.

\bibitem[Sengupta and Kambhampati(2020)]{sengupta2020multi}
Sailik Sengupta and Subbarao Kambhampati.
\newblock Multi-agent reinforcement learning in bayesian stackelberg markov games for adaptive moving target defense.
\newblock \emph{arXiv preprint arXiv:2007.10457}, 2020.

\bibitem[Zhuang et~al.(2014)Zhuang, DeLoach, and Ou]{zhuang2014towards}
Rui Zhuang, Scott~A DeLoach, and Xinming Ou.
\newblock Towards a theory of moving target defense.
\newblock In \emph{Proceedings of the first ACM workshop on moving target defense}, pages 31--40, 2014.

\bibitem[Jajodia et~al.(2012)Jajodia, Ghosh, Subrahmanian, Swarup, Wang, and Wang]{jajodia2012moving}
Sushil Jajodia, Anup~K Ghosh, VS~Subrahmanian, Vipin Swarup, Cliff Wang, and X~Sean Wang.
\newblock \emph{Moving Target Defense II: Application of Game Theory and Adversarial Modeling}, volume 100.
\newblock Springer, 2012.

\bibitem[Zhu and Ba{\c{s}}ar(2013)]{zhu2013game}
Quanyan Zhu and Tamer Ba{\c{s}}ar.
\newblock Game-theoretic approach to feedback-driven multi-stage moving target defense.
\newblock In \emph{International conference on decision and game theory for security}, pages 246--263. Springer, 2013.

\bibitem[Eghtesad et~al.(2020)Eghtesad, Vorobeychik, and Laszka]{eghtesad2020adversarial}
Taha Eghtesad, Yevgeniy Vorobeychik, and Aron Laszka.
\newblock Adversarial deep reinforcement learning based adaptive moving target defense.
\newblock In \emph{Decision and Game Theory for Security: 11th International Conference, GameSec 2020, College Park, MD, USA, October 28--30, 2020, Proceedings 11}, pages 58--79. Springer, 2020.

\bibitem[Xu et~al.(2021)Xu, Zhang, Kuang, Zhou, and Yu]{xu2021context}
Changqiao Xu, Tao Zhang, Xiaohui Kuang, Zan Zhou, and Shui Yu.
\newblock Context-aware adaptive route mutation scheme: A reinforcement learning approach.
\newblock \emph{IEEE Internet of Things Journal}, 8\penalty0 (17):\penalty0 13528--13541, 2021.

\bibitem[Li and Zheng(2023)]{meta-rl}
Henger Li and Zizhan Zheng.
\newblock Robust moving target defense against unknown attacks: A meta-reinforcement learning approach.
\newblock page 107–126, Berlin, Heidelberg, 2023. Springer-Verlag.
\newblock ISBN 978-3-031-26368-2.
\newblock \doi{10.1007/978-3-031-26369-9_6}.
\newblock URL \url{https://doi.org/10.1007/978-3-031-26369-9_6}.

\bibitem[Zheng and Siami~Namin(2019)]{8754059}
Jianjun Zheng and Akbar Siami~Namin.
\newblock Enforcing optimal moving target defense policies.
\newblock In \emph{2019 IEEE 43rd Annual Computer Software and Applications Conference (COMPSAC)}, volume~1, pages 753--759, 2019.
\newblock \doi{10.1109/COMPSAC.2019.00112}.

\bibitem[Liu et~al.(2018)Liu, Wu, Pahwa, Ding, Ibrahim, and Liu]{liu2018hidden}
Bo~Liu, Hongyu Wu, Anil Pahwa, Fei Ding, Erfan Ibrahim, and Ting Liu.
\newblock Hidden moving target defense against false data injection in distribution network reconfiguration.
\newblock In \emph{2018 IEEE Power \& Energy Society General Meeting (PESGM)}, pages 1--5. IEEE, 2018.

\bibitem[Makanju et~al.(2017)Makanju, Zincir-Heywood, and Kiyomoto]{makanju2017evolutionary}
Adetokunbo Makanju, A~Nur Zincir-Heywood, and Shinsaku Kiyomoto.
\newblock On evolutionary computation for moving target defense in software defined networks.
\newblock In \emph{Proceedings of the genetic and evolutionary computation conference companion}, pages 287--288, 2017.

\bibitem[Kamhoua et~al.(2021)Kamhoua, Kiekintveld, Fang, and Zhu]{kamhoua2021game}
Charles~A Kamhoua, Christopher~D Kiekintveld, Fei Fang, and Quanyan Zhu.
\newblock \emph{Game theory and machine learning for cyber security}.
\newblock John Wiley \& Sons, 2021.

\bibitem[Puterman(1990)]{puterman1990markov}
Martin~L Puterman.
\newblock Markov decision processes.
\newblock \emph{Handbooks in operations research and management science}, 2:\penalty0 331--434, 1990.

\bibitem[S{\'a}nchez et~al.(2021)S{\'a}nchez, Valero, Celdr{\'a}n, Bovet, P{\'e}rez, and P{\'e}rez]{sanchez2021survey}
Pedro Miguel~S{\'a}nchez S{\'a}nchez, Jose Maria~Jorquera Valero, Alberto~Huertas Celdr{\'a}n, G{\'e}r{\^o}me Bovet, Manuel~Gil P{\'e}rez, and Gregorio~Mart{\'\i}nez P{\'e}rez.
\newblock A survey on device behavior fingerprinting: Data sources, techniques, application scenarios, and datasets.
\newblock \emph{IEEE Communications Surveys \& Tutorials}, 23\penalty0 (2):\penalty0 1048--1077, 2021.

\bibitem[Zhang et~al.(2022)Zhang, Ning, Shi, Farha, Xu, Xu, Zhang, and Choo]{zhang2022artificial}
Zhimin Zhang, Huansheng Ning, Feifei Shi, Fadi Farha, Yang Xu, Jiabo Xu, Fan Zhang, and Kim-Kwang~Raymond Choo.
\newblock Artificial intelligence in cyber security: research advances, challenges, and opportunities.
\newblock \emph{Artificial Intelligence Review}, pages 1--25, 2022.

\bibitem[Guestrin et~al.(2003)Guestrin, Koller, Parr, and Venkataraman]{guestrin2003efficient}
Carlos Guestrin, Daphne Koller, Ronald Parr, and Shobha Venkataraman.
\newblock Efficient solution algorithms for factored mdps.
\newblock \emph{Journal of Artificial Intelligence Research}, 19:\penalty0 399--468, 2003.

\bibitem[Murphy(2002)]{murphy2002dynamic}
Kevin~Patrick Murphy.
\newblock \emph{Dynamic bayesian networks: representation, inference and learning}.
\newblock University of California, Berkeley, 2002.

\bibitem[Paruchuri et~al.(2008)Paruchuri, Pearce, Marecki, Tambe, Ordonez, and Kraus]{dobss}
Praveen Paruchuri, Jonathan~P. Pearce, Janusz Marecki, Milind Tambe, Fernando Ordonez, and Sarit Kraus.
\newblock Playing games for security: An efficient exact algorithm for solving bayesian stackelberg games.
\newblock AAMAS '08, page 895–902, Richland, SC, 2008. International Foundation for Autonomous Agents and Multiagent Systems.
\newblock ISBN 9780981738116.

\bibitem[Kiekintveld et~al.(2009)Kiekintveld, Jain, Tsai, Pita, Ord{\'o}nez, and Tambe]{kiekintveld2009computing}
Christopher Kiekintveld, Manish Jain, Jason Tsai, James Pita, Fernando Ord{\'o}nez, and Milind Tambe.
\newblock Computing optimal randomized resource allocations for massive security games.
\newblock 2009.

\bibitem[Breton et~al.(1988)Breton, Alj, and Haurie]{breton1988sequential}
Michele Breton, Abderrahmane Alj, and Alain Haurie.
\newblock Sequential stackelberg equilibria in two-person games.
\newblock \emph{Journal of Optimization Theory and Applications}, 59:\penalty0 71--97, 1988.

\bibitem[Lei et~al.(2017)Lei, Ma, and Zhang]{lei2017optimal}
Cheng Lei, Duo-He Ma, and Hong-Qi Zhang.
\newblock Optimal strategy selection for moving target defense based on markov game.
\newblock \emph{IEEE Access}, 5:\penalty0 156--169, 2017.

\bibitem[Chowdhary et~al.(2019)Chowdhary, Sengupta, Alshamrani, Huang, and Sabur]{chowdhary2019adaptive}
Ankur Chowdhary, Sailik Sengupta, Adel Alshamrani, Dijiang Huang, and Abdulhakim Sabur.
\newblock Adaptive mtd security using markov game modeling.
\newblock In \emph{2019 International Conference on Computing, Networking and Communications (ICNC)}, pages 577--581. IEEE, 2019.

\bibitem[Viswanathan et~al.(2022)Viswanathan, Bose, and Paruchuri]{viswanathan2022moving}
Vignesh Viswanathan, Megha Bose, and Praveen Paruchuri.
\newblock Moving target defense under uncertainty for web applications.
\newblock In \emph{Proceedings of the 21st International Conference on Autonomous Agents and Multiagent Systems}, pages 1750--1752, 2022.

\bibitem[Arora et~al.(2012)Arora, Dekel, and Tewari]{policy-regret}
Raman Arora, Ofer Dekel, and Ambuj Tewari.
\newblock Online bandit learning against an adaptive adversary: from regret to policy regret.
\newblock In \emph{Proceedings of the 29th International Coference on International Conference on Machine Learning}, ICML'12, page 1747–1754, Madison, WI, USA, 2012. Omnipress.
\newblock ISBN 9781450312851.

\bibitem[Mladenov et~al.()Mladenov, Boutilier, Schuurmans, Meshi, Elidan, and Lu]{mladenovlogistic}
Martin Mladenov, Craig Boutilier, Dale Schuurmans, Ofer Meshi, Gal Elidan, and Tyler Lu.
\newblock Logistic markov decision processes.

\bibitem[Li et~al.(2020)Li, Shen, and Zheng]{li2020spatial}
Henger Li, Wen Shen, and Zizhan Zheng.
\newblock Spatial-temporal moving target defense: A markov stackelberg game model.
\newblock \emph{arXiv preprint arXiv:2002.10390}, 2020.

\bibitem[Team.(2021)]{msft:cyberbattlesim}
Microsoft Defender~Research Team.
\newblock Cyberbattlesim.
\newblock \url{https://github.com/microsoft/cyberbattlesim}, 2021.
\newblock Created by Christian Seifert, Michael Betser, William Blum, James Bono, Kate Farris, Emily Goren, Justin Grana, Kristian Holsheimer, Brandon Marken, Joshua Neil, Nicole Nichols, Jugal Parikh, Haoran Wei.

\bibitem[Mladenov et~al.(2017)Mladenov, Boutilier, Schuurmans, Elidan, Meshi, and Lu]{mladenov2017approximate}
Martin Mladenov, Craig Boutilier, Dale Schuurmans, Gal Elidan, Ofer Meshi, and Tyler Lu.
\newblock Approximate linear programming for logistic markov decision processes.
\newblock 2017.

\bibitem[Liu and Su(2020)]{liu2020regret}
Shuang Liu and Hao Su.
\newblock Regret bounds for discounted mdps.
\newblock \emph{arXiv preprint arXiv:2002.05138}, 2020.

\bibitem[Zhou et~al.(2021)Zhou, He, and Gu]{zhou2021provablyregret}
Dongruo Zhou, Jiafan He, and Quanquan Gu.
\newblock Provably efficient reinforcement learning for discounted mdps with feature mapping.
\newblock In \emph{International Conference on Machine Learning}, pages 12793--12802. PMLR, 2021.

\bibitem[He et~al.(2021)He, Zhou, and Gu]{he2021nearlyregret}
Jiafan He, Dongruo Zhou, and Quanquan Gu.
\newblock Nearly minimax optimal reinforcement learning for discounted mdps.
\newblock \emph{Advances in Neural Information Processing Systems}, 34:\penalty0 22288--22300, 2021.

\end{thebibliography}


\appendix
\newpage
\onecolumn

\section{Approximate Linear Program Formulation}
\label{appen:alp}
The primal LP formulation \citep{puterman1990markov} to solve for the optimal policy in a finite MDP goes as follows:

\begin{equation}
\min _{v_{s}: s \in S} \sum_{s \in S} \theta_{s} v_{s}
\end{equation}

\begin{equation}\label{eq:primal-lp-const}
    \text { s.t. } 0 \geq Q^{v}(s, a) - v_{s}, \forall s \in S, a \in A
\end{equation}

where $Q^v(\cdot, a)$ denotes the action-value backup, $v_s$ denotes the optimal value function at $s$ and $\theta$ denotes the initial state distribution. In the factored MDP model, the value function is approximated using a linear combination of basis functions denoted as $\mathcal{B} = \{\beta_1, \ldots, \beta_k\}$ over state variables $X$. Each $\beta_i$ is a function of a small set $B_i \subset \mathcal{S}$, including a bias factor $\beta_{\text{bias}} = 1$ in $\mathcal{B}$ and $\operatorname{Par}_{B_{i}}= \cup_{S^{j} \in B_{i}} \operatorname{Par}_{j}$ \citep{guestrin2003efficient}. The value function is parameterized by a weight vector $\mathbf{w}$:

\[
V(\boldsymbol{s}; \mathbf{w}) = \sum_{i \leq k} w_i \beta_i(\boldsymbol{s}[B_i]) = \mathbf{B}_{\boldsymbol{s}} \mathbf{w},
\]

where $\mathbf{B} = [\beta_1, \ldots, \beta_k]$ is the basis matrix, and $\mathbf{B}_{\boldsymbol{s}}$ denotes the row of $\mathbf{B}$ corresponding to state $\boldsymbol{s}$. When $\theta$ is factored similarly and $g_i$ denotes the backprojection \cite{guestrin2003efficient} of basis function $\beta_i$, the primal LP formulation in the MTD scenario can be expressed as:

\begin{equation}
\begin{aligned}
\min _{\mathbf{w}} & \sum_{i \leq k} \sum_{\mathbf{b}_{i} \in \operatorname{Dom}\left(B_{i}\right)} \theta\left[\mathbf{b}_{i}\right] w_{i} \beta_{i}\left(\mathbf{b}_{i}\right) \\\\
\text { s.t. } 
& 0 \geq C(\mathbf{s}, \mathbf{a} ; \mathbf{w}) \quad \forall \, \mathbf{s}, \mathbf{a} \quad \text { where } \\
& C(\mathbf{s}, \mathbf{a} ; \mathbf{w})=\sum_{\varphi \in\{0, 1\}} c(\mathbf{s}, \mathbf{a}, \varphi ; \mathbf{w}) .
\end{aligned}
\end{equation}

For $\varphi = 1$,

\begin{equation}
\begin{aligned}
& c(\mathbf{s}, \mathbf{a}, \varphi ; \mathbf{w})  \\
& = \sum_{\tau} \operatorname{P}(\varphi\mid\mathbf{s}, \mathbf{a}, \tau) \operatorname{\hat{P}_{att}}(\tau\mid\mathbf{s}, \mathbf{a}) \big[M - l(\tau, (\mathbf{s}, \mathbf{a})) - \alpha \cdot sc(\mathbf{s}, \mathbf{a}) + \sum_{i \leq k} w_{i}\left(\gamma \cdot g_{i}\left(\mathbf{s}\left[\operatorname{Par}_{B_{i}}\right], \mathbf{a}\left[\operatorname{Par}_{B_{i}}\right]\right)-\beta_{i}\left(\mathbf{s}\left[B_{i}\right]\right)\right)\big]\\
& = \sum_{\tau} \mu(\tau, (\mathbf{s}, \mathbf{a})) \operatorname{\hat{P}_{att}}(\tau\mid\mathbf{s}, \mathbf{a}) \big[M -l(\tau, (\mathbf{s}, \mathbf{a})) - \alpha \cdot sc(\mathbf{s}, \mathbf{a}) + \sum_{i \leq k} w_{i}\left(\gamma \cdot g_{i}\left(\mathbf{s}\left[\operatorname{Par}_{B_{i}}\right], \mathbf{a}\left[\operatorname{Par}_{B_{i}}\right]\right)-\beta_{i}\left(\mathbf{s}\left[B_{i}\right]\right)\right)\big]
\end{aligned}
\end{equation}

\begin{equation}
\begin{aligned}
\label{eq:alp-constr1}
& = \sum_{\tau} \mu(\tau, (\mathbf{s}, \mathbf{a})) \operatorname{\hat{P}_{att}}(\tau\mid\mathbf{s}, \mathbf{a}) \big[M -l(\tau, (\mathbf{s}, \mathbf{a})) - \alpha \cdot sc(\mathbf{s}, \mathbf{a}) + \sum_{i \leq k} w_{i}\gamma \cdot g_{i}\left(\mathbf{s}\left[\operatorname{Par}_{B_{i}}\right], \mathbf{a}\left[\operatorname{Par}_{B_{i}}\right]\right) \big]\\
& - \sum_{\tau} \mu(\tau, (\mathbf{s}, \mathbf{a})) \operatorname{\hat{P}_{att}}(\tau\mid\mathbf{s}, \mathbf{a})\big[\sum_{i \leq k} w_{i}\beta_{i}\left(\mathbf{s}\left[B_{i}\right]\right) \big]
\end{aligned}
\end{equation}

as $\mu(\tau, (\mathbf{s}, \mathbf{a}))$ is the attack success rate of $\tau$ given $(\mathbf{s}, \mathbf{a})$. One can see that the constraint Eq. \ref{eq:alp-constr1} is the difference between the action-value backup and the value function as seen in Eq. \ref{eq:primal-lp-const}. Similarly, the constraints for $\varphi = 0$ are given below. However, here, since loss from attack is $0$ and switching cost doesn't depend on the attacker response, we have

\begin{equation}
\begin{aligned}
\label{eq:alp-constr2}
& c(\mathbf{s}, \mathbf{a}, \varphi ; \mathbf{w}) = \operatorname{P}(\varphi\mid\mathbf{s}, \mathbf{a})  \big[M - \alpha \cdot sc(\mathbf{s}, \mathbf{a}) + \sum_{i \leq k} w_{i}\left(\gamma \cdot g_{i}\left(\mathbf{s}\left[\operatorname{Par}_{B_{i}}\right], \mathbf{a}\left[\operatorname{Par}_{B_{i}}\right]\right)-\beta_{i}\left(\mathbf{s}\left[B_{i}\right]\right)\right)\big]
\end{aligned}
\end{equation}

Notice that, 

\begin{equation}
\begin{aligned}
    & \operatorname{P}(\varphi=0\mid\mathbf{s}, \mathbf{a}) = 1 - \operatorname{P}(\varphi=1\mid\mathbf{s}, \mathbf{a}) = 1 - \sum_{\tau}\operatorname{P}(\varphi=1\mid\mathbf{s}, \mathbf{a}, \tau) \operatorname{\hat{P}_{att}}(\tau\mid\mathbf{s}, \mathbf{a}) 
\end{aligned}
\end{equation}
\begin{equation}
\begin{aligned}
    = 1 - \sum_{\tau} \mu(\tau, (\mathbf{s}, \mathbf{a})) \operatorname{\hat{P}_{att}}(\tau\mid\mathbf{s}, \mathbf{a})
\end{aligned}
\end{equation}

Here, $\boldsymbol{s}[U]$ (similarly $\boldsymbol{a}[U]$) denotes the restriction of a variable instantiation to the subset $U \in \mathcal{S} \cup \mathcal{A}$. Following \citep{mladenov2017approximate}, we use sparse binarized features in our experiments for states and actions, along with corresponding indicator basis functions in solving the ALP.

\section{Attacker Type Probability Estimation}
\label{appen:est}
Let the success of an attack executed by attacker type $\tau$ for each state-action pair $(s, a)$ be characterized by independent Bernoulli random variables with parameter $p_{\tau, s, a}$ over $T$ timesteps. We denote them by $\{X_1^{\tau, s, a}, X_2^{\tau, s, a}, ..., X_T^{\tau, s, a}\}$. Let us define random variable $N^{\tau, s, a}$ as follows:

\[Y^{\tau, s, a} = \sum_{t=1}^{T}\frac{1}{\beta^{T-t}}X_t^{\tau, s, a}\]

where $\beta$ is the hyper-parameter from $ATA-FMDP$. Taking expectation of $N^{\tau, s, a}$, we get

\[ \mathbb{E}[N^{\tau, s, a}] = \mathbb{E}[\sum_{t=1}^{T}\frac{1}{\beta^{T-t}}X_t^{\tau, s, a}] = \sum_{t=1}^{T}\frac{1}{\beta^{T-t}}\mathbb{E}[X_t^{\tau, s, a}] = \sum_{t=1}^{T}\frac{1}{\beta^{T-t}}p_{\tau, s, a} \]

We can write $p_{\tau, s, a}$ as $P(\varphi=1 | \tau, s, a) P(\tau | s, a) P(s, a)$ as it is the joint probability of an attack success when the defender is in state $s$, takes action $a$ and is attacked by attacker type $\tau$. Here, $P(\varphi=1 | \tau, s, a)$ is the attack success rate $\mu(\tau, (s, a))$ of attacker type $\tau$ for state-action pair $(s, a)$. We aim to estimate $P(\tau | s, a)$ or $P_{att}(\tau | s, a)$. Let us assume that the true value of $P(\tau | s, a)$ remains fixed in these $T$ timesteps. Hence, we have

\[ \mathbb{E}[N^{\tau, s, a}] = \sum_{t=1}^{T}\frac{1}{\beta^{T-t}}p_{\tau, s, a}  = \sum_{t=1}^{T}\frac{1}{\beta^{T-t}}P(s, a)P_{att}(\tau | s, a) \mu(\tau, (s, a))\]
\[=\left(\frac{1-\frac{1}{\beta^T}}{1-\frac{1}{\beta}}\right)P(s, a)P_{att}(\tau | s, a) \mu(\tau, (s, a))\]

\[\mathbb{E}\left[\frac{N^{\tau, s, a}}{\mu(\tau, (s, a))}\right] = \left(\frac{1-\frac{1}{\beta^T}}{1-\frac{1}{\beta}}\right)P(s, a)P_{att}(\tau | s, a) \]

From above, we that value of $\frac{N^{\tau, s, a}}{\mu(\tau, (s, a))}$ acts as an unbiased estimator of a constant $\left(\frac{1-\frac{1}{\beta^T}}{1-\frac{1}{\beta}}\right)$ times $P(\tau, s, a)$. The value taken by $N^{\tau, s, a}$ is calculated as $n_{\tau, s, a}$ in $ATA-FMDP$. For state-action pair $(s, a)$, estimate $\hat{P}_{att}$ is calculated from $\frac{n_{\tau, s, a}}{\mu(\tau, (s, a))}$ values normalized over the attacker types.

\section{Regret Analysis}

To argue whether an algorithm performs well across different problem instances, a standard approach is to compare how "badly" the chosen actions perform compared to the scenario where the optimal action is chosen in each timestep. Hence, regret is a quantity defined to quantify the "regret" an algorithm has for choosing the actions taken compared to the maximum that it could have achieved if it had known the best action in advance. Regret can take various forms, depending on the context in which it is applied. 

\subsection{Proof of Theorem \ref{thm:policy-regret}}
\label{appen:policy-regret}
\textit{For any MTD defense strategy on $n>1$ configurations, $\exists$ an adaptive adversary such that the defender's policy regret compared to the best static configuration in hindsight is $\Omega(T)$.}

\begin{proof}

Using the definition of policy regret from \cite{policy-regret}, we have
\begin{equation}
    \max_{\left(b_1, b_2, \ldots, b_T\right) \in \mathcal{C}_T} \sum_{t=1}^T r_t(b_1, b_2, \ldots, b_t)-\mathbb{E}[\sum_{t=1}^T r_t(\mathbf{A}_{1, \ldots, t})]
\end{equation}

where random variable $\mathbf{A}_i$ represents the action taken at the $i^{th}$ timestep. Since we are comparing against the best constant action in hindsight, the competitor class $\mathcal{C}_T$ consists of sequences of the form $(b,b,..., b)$. The policy regret hence becomes

\begin{equation}
    \max_{\left(b, b, \ldots, b\right) \in \mathcal{C}_T} \sum_{t=1}^T r_t(b, b, \ldots, b)-\mathbb{E}[\sum_{t=1}^T r_t(\mathbf{A}_{1, \ldots, t})]
\end{equation}

Considering switching costs, we observe that constant sequences of competitor class suffer from no switching cost. However, the switching cost needs to be subtracted from the algorithm's rewards, giving us the policy regret

\begin{equation}
\label{eq:policy-regret-sc}
    \max_{\left(b, b, \ldots, b\right) \in \mathcal{C}_T} \sum_{t=1}^T r_t(b, b, \ldots, b)-\mathbb{E}[\sum_{t=1}^T r_t(\mathbf{A}_{1, \ldots, t}) - sc(\mathbf{A}_{t-1}, \mathbf{A}_t)]
\end{equation}

Let us define an adaptive adversary that chooses the reward function as follows:
\begin{equation}
    r_t(s_1, s_2, \cdots, s_t) = 
    \begin{cases}
        0 & \text{if } s_1 = y \text{ and } t > 1, \\
        1 & otherwise.
    \end{cases}
\end{equation}

Considering $T+1$ timesteps, we have

\begin{equation}
    \max_{\left(b, b, \ldots, b\right) \in \mathcal{C}_{T+1}} \sum_{t=1}^{T+1} r_t(b, b, \ldots, b) = \sum_{t=1}^{T+1} r_t(x, x, \ldots, x) = T+1
\end{equation}

where $x \neq y$. Using this value in (\ref{eq:policy-regret-sc})., we get policy regret of

\begin{equation}
\begin{aligned}
    T+1 -\mathbb{E}[\sum_{t=1}^{T+1} r_t(\mathbf{A}_{1, \ldots, t}) - sc(\mathbf{A}_{t-1}, \mathbf{A}_t)] &= T+1 + \mathbb{E}[\sum_{t=1}^{T+1} -r_t(\mathbf{A}_{1, \ldots, t}) + sc(\mathbf{A}_{t-1}, \mathbf{A}_t)]
\end{aligned}
\end{equation}

Let us look at the quantity within expectation for a sequence of configurations $(s_0, s_1, s_2, \cdots, s_{T+1})$. Given switching cost, $sc(s, s')=0$ if $s=s'$ and $\in (0, 1]$ if $s \neq s'$, using definition of $r_t(s_1, s_2, \cdots, s_t)$, the quantity at timestep $t > 1$ is

\[
- r_t(s_1, s_2, \cdots, s_t) + sc(s_{t-1}, s_t) = 
\begin{cases}
    sc(s_{t-1}, s_t) & \text{if } s_1=y \text{ and } s_{t-1} \neq s_t, \\
    0 & \text{if } s_1=y \text{ and } s_{t-1}=s_t, \\
    -1 + sc(s_{t-1}, s_t) & \text{if } s_1 \neq y \text{ and } s_{t-1} \neq s_t \\
    -1 & \text{if } s_1 \neq y \text{ and } s_{t-1}=s_t.
\end{cases}
\]

and for $t = 1$,

\[
- r_t(s_1, s_2, \cdots, s_t) + sc(s_{t-1}, s_t) = 
\begin{cases}
    -1 + sc(s_{t-1}, s_t) & \text{if } s_{t-1} \neq s_t \\
    -1 & \text{if } s_{t-1}=s_t.
\end{cases}
\]

From the above equations, we have for $t>1$ and $s_1 = y$,

\[
- r_t(s_1, s_2, \cdots, s_t) + sc(s_{t-1}, s_t) \geq 0
\]

For $t>1$ and $s_1 \neq y$,

\[
- r_t(s_1, s_2, \cdots, s_t) + sc(s_{t-1}, s_t) \geq -1
\]

And, for $t=1$, $- r_t(s_1, s_2, \cdots, s_t) + sc(s_{t-1}, s_t) \geq -1$. Let Pr$[s_1=y] = p$. Taking summation over $T+1$ timesteps, considering the expected value, we have policy regret

\begin{equation}
    T+1+\mathbb{E}[\sum_{t=1}^{T+1} -r_t(\mathbf{A}_{1, \ldots, t}) + sc(\mathbf{A}_{t-1}, \mathbf{A}_t)] \geq T+1 - 1 + pT0 +(1-p)T(-1) = pT
\end{equation}

Hence, we can not obtain a sublinear regret in this case.

\end{proof}

\subsection{Regret Analysis on Discounted MDP}
\label{appen:regret}
In the MTD problem, the defender follows a policy while the adaptive attacker responds to it. Let us define the true attacker policy as $\nu_{true}^{\pi_d}: \mathcal{S} \rightarrow \Delta (\mathcal{T})$ where the attacker responds to the defender policy $\pi_d$ and $\mathcal{T}$ is the set of all attacker types. $\nu_{true}^{\pi_d}(s)$ decides the probability of each attacker type attacking state $s$. Let us see how this policy connects to the attacker-type probabilities ($\hat{P}_{att}$) we estimate in Algorithm \ref{algorithm:ATA-MDP-algo}. Notice that as in each timestep the defender and attacker take actions independently without knowledge of the opponent's action, for $\mathcal{T}=\{\tau_1, \tau_2, ... \tau_k\}$,

$$P(\tau_i|s, a) = \frac{\nu_{true}^{\pi_d}(\tau_i|s) \pi_d(a|s)}{\sum_{\tau \in \mathcal{T}}\nu_{true}^{\pi_d}(\tau|s)\pi_d(a|s)}$$

These are the true underlying attacker-type probabilities, but the defender has access to only an estimate of these probabilities, i.e., $\hat{P}_{att}$ (or simply denoted by $\hat{P}(\tau|s,a))$. The attack loss, as defined in the defender reward, is:

$$al(s, a) = \sum_{\tau \in \mathcal{T}} P(\tau|s, a) \mu(\tau, (s, a)) l(\tau, (s, a))$$

Let us denote the attack loss calculated using the estimated attacker-type probabilities by $\tilde{al}$:

$$\tilde{al}(s, a) = \sum_{\tau \in \mathcal{T}} \hat{P}(\tau|s, a) \mu(\tau, (s, a)) l(\tau, (s, a))$$

Hence, the true defender rewards amount to $r_s^a=M-al(s,a)-sc(s,a) \,\forall s,a$ while the rewards fed into the factored MDP for optimal policy calculation are $M-\tilde{al}(s,a)-sc(s,a) \,\forall s,a$. Let us denote these rewards by $\tilde{r}_a^s$.

In Algorithm \ref{algorithm:ATA-MDP-algo}, we re-calculate the MDP policy based on the latest attacker-type probability estimates when triggered. These triggers can be defined by the defender based on attacks observed or system requirements. The idea is to ensure that we have the optimal defender policy against the attacker behavior observed till now until the attacker behavior changes substantially, which would render the defender policy calculated obsolete. 

Let us consider that the re-optimization occurs at timesteps $\{0, T_1, T_2, ..., T_k\}$ in the time period $[0, T)$, $T_0=0$, and $T_{k+1}=T$. Notice that within each interval $[T_i, T_{i+1})$ for $i \in [k] \cup \{0\}$, the defender policy remains the same. Hence, let us characterize the intervals based on the defender policy adopted in that interval. For instance, if the defender follows the policy $\pi_d$ in $[T_i, T_{i+1})$, we call the interval a $\pi_d$-period. 

Let $[T_i, T_{i+1})$ be a $\pi_d$-period. The adaptive attacker responds to the defender policy with $\nu_{true}^{\pi_d}$. This may or may not be the best response to the defender policy. No assumption is made regarding the quality of this attacker policy. We assume that within each of these intervals, the attacker policy doesn't change abruptly. The attacker policy directly dictates the attacker-type probability estimates, which in turn dictates the defender reward. Hence, we assume that within these intervals, the difference between true defender rewards and estimated defender rewards is bounded, i.e.,  $|r_s^a - \tilde{r}_s^a| < \epsilon$ where $\epsilon > 0$. Now, let $V^{*, \nu_{true}^{\pi_d}}$ be the optimal value function that the factored MDP would have calculated if it had used the true defender rewards derived from $\nu_{true}^{\pi_d}$ instead of the estimates and $V^{\pi_d, \nu_{est}}$ be the optimal value function based on the optimal policy $\pi_d$ calculated using the estimated defender rewards corresponding to an underlying estimated attacker policy $\nu_{est}$. These optimal value functions satisfy the Bellman equations as follows:

$$V^{*, \nu_{true}^{\pi_d}}(s) = \max_{a \in \mathcal{A}} (r_s^a + \gamma V^{*, \nu_{true}^{\pi_d}}(s'))$$

where $s'$ is determined deterministically from state $s$ on taking action $a$. Hence, we replace $s'$ with the tuple $(s, a)$ and write $V^{*, \nu_{true}^{\pi_d}}$ as follows:

$$V^{*, \nu_{true}^{\pi_d}}(s) = \max_{a \in \mathcal{A}} (r_s^a + \gamma V^{*, \nu_{true}^{\pi_d}}((s, a)))$$

Let $f(s,a) = r_s^a + \gamma V^{*, \nu_{true}^{\pi_d}}((s, a))$. Hence, $V^{*, \nu_{true}^{\pi_d}}(s) = \max_{a \in \mathcal{A}} f(s,a)$. Similarly, the optimal value function calculated from the estimated defender rewards corresponding to the attacker policy, $\nu_{est}$, satisfies the following Bellman equation:

$$V^{*, \nu_{est}}(s) = V^{\pi_d, \nu_{est}}(s) = \max_{a \in \mathcal{A}} (\tilde{r}_s^a + \gamma V^{\pi_d, \nu_{est}}((s, a)))$$

Let $g(s,a) = \tilde{r}_s^a + \gamma V^{\pi_d, \nu_{est}}((s, a))$. Hence, $V^{\pi_d, \nu_{est}}(s) = \max_{a \in \mathcal{A}} g(s,a)$.

We know, $\forall s,a$
\[f(s,a) - g(s,a) \leq |f(s,a) - g(s,a)|\iff f(s,a) \leq |f(s,a) - g(s,a)| + g(s,a)\]

\[\therefore \max_{a \in \mathcal{A}}f(s,a) \leq \max_{a \in \mathcal{A}} (|f(s,a) - g(s,a)| + g(s,a))\leq \max_{a \in \mathcal{A}} |f(s,a) - g(s,a)| + \max_{a \in \mathcal{A}} g(s,a)\]

Hence, \[\max_{a \in \mathcal{A}}f(s,a) - \max_{a \in \mathcal{A}}g(s,a) \leq \max_{a \in \mathcal{A}}|f(s,a) - g(s,a)|\]. 

Similarly, $\max_{a \in \mathcal{A}}g(s,a) - \max_{a \in \mathcal{A}}f(s,a) \leq \max_{a \in \mathcal{A}}|g(s,a) - f(s,a)|$. Hence, we have

\[|\max_{a \in \mathcal{A}}f(s,a) - \max_{a \in \mathcal{A}}g(s,a)| \leq \max_{a \in \mathcal{A}}|f(s,a) - g(s,a)|\]

Substituting $f(s,a)$ and $g(s,a)$ and using $|r_s^a - \tilde{r}_s^a| < \epsilon$, we get

\[|\max_{a \in \mathcal{A}}(r_s^a + \gamma V^{*, \nu_{true}^{\pi_d}} ((s, a))) - \max_{a \in \mathcal{A}}(\tilde{r}_s^a + \gamma V^{\pi_d, \nu_{est}}((s, a)))| \leq \max_{a \in \mathcal{A}}|r_s^a + \gamma V^{*, \nu_{true}^{\pi_d}}((s, a)) - \tilde{r}_s^a - \gamma V^{\pi_d, \nu_{est}}((s, a))|\]
\[\leq \max_{a \in \mathcal{A}}|r_s^a - \tilde{r}_s^a| + \gamma |V^{*, \nu_{true}^{\pi_d}}((s, a))-V^{\pi_d, \nu_{est}}((s, a))|\]
\[\leq \epsilon + \gamma \max_{a \in \mathcal{A}}|V^{*, \nu_{true}^{\pi_d}}((s, a))-V^{\pi_d, \nu_{est}}((s, a))|\]

Hence,

\[|\max_{a \in \mathcal{A}}(r_s^a + \gamma V^{*, \nu_{true}^{\pi_d}}((s, a))) - \max_{a \in \mathcal{A}}(\tilde{r}_s^a + \gamma V^{\pi_d, \nu_{est}}((s, a)))| \leq \epsilon + \gamma \max_{a \in \mathcal{A}}|V^{*, \nu_{true}^{\pi_d}}((s, a))-V^{\pi_d, \nu_{est}}((s, a))|\]

\[|V^{*, \nu_{true}^{\pi_d}}(s) - V^{\pi_d, \nu_{est}}(s)| \leq \epsilon + \gamma \max_{a \in \mathcal{A}}|V^{*, \nu_{true}^{\pi_d}}((s, a))-V^{\pi_d, \nu_{est}}((s, a))|\]

Since $(s,a)$ represents the next state reached on taking action $a$ in state $s$, we can recurse the bound on $|V^{*, \nu_{true}^{\pi_d}}((s, a))-V^{\pi_d, \nu_{est}}((s, a))|$ from $T_i$ to $T_{i+1}$. However, $T_{i+1}$ is not known at $T_i$ when the optimal policy is re-calculated. So, at timestep $T_i$ when the optimal policy is calculated based on estimated reward structure $\tilde{r}_s^a$, the problem at hand is to solve the infinite horizon factored MDP for the optimal policy to be adopted in $[T_i,\infty)$. So, we have

\begin{equation}
\label{eq:val-diff}
    |V^{*, \nu_{true}^{\pi_d}}(s) - V^{\pi_d, \nu_{est}}(s)| \leq \epsilon + \gamma \epsilon + \gamma^2 \epsilon + ... = \frac{\epsilon}{1-\gamma}
\end{equation}

Therefore, if $\pi_d$ is the optimal policy calculated at $T_i$, as long as the maximum error in reward estimates $\max_{s, a} |r_s^a-\tilde{r}^s_a| < (1-\gamma)\epsilon$, $\epsilon > 0$ and $\gamma \in [0, 1)$, the maximum error in optimal value function remains $|V^{*, \nu_{true}^{\pi_d}}(s) - V^{\pi_d, \nu_{est}}(s)| < \epsilon$.

The standard metric to evaluate the performance of an agent over a finite number of $T$ timesteps is the cumulative regret with respect to the sequence of stationary policies {$\pi_t$} at each timestep learned by the algorithm. However, here, we look at the interval $[T_i, T_{i+1})$ and within this interval, the defender agent follows $\pi_d$. 

For discounted infinite horizon MDPs, the commonly used definition of regret is the cumulative sub-optimality in the value function of the chosen policy compared to the optimal policy. 
\[
\text{Regret} = \sum_{t=1}^T \left( V^*(s_t) - V^{\pi}(s_t) \right),
\]
Previous works \cite{liu2020regret, zhou2021provablyregret, he2021nearlyregret} have discussed and motivated this notion of regret, relating it to the "sample complexity of exploration" in discounted MDPs. Building on this notion of regret, we define the average regret for time interval $[T_i, T_{i+1}]$ in our case as follows:

$$AvgRegret(T_i, T_{i+1}) = \frac{1}{T_{i+1}-T_i} \sum_{t=T_i}^{T_{i+1}-1}(V^{*, \nu_{true}^{\pi_d}}(s_t)-V^{\pi_d, \nu_{true}^{\pi_d}}(s_t))$$

Notice that even though $\pi_d$ is the optimal policy against $\nu_{est}$, it may not be optimal against the true attacker policy $\nu_{true}^{\pi_d}$. From Eq. \ref{eq:val-diff}, we have

\[|V^{*, \nu_{true}^{\pi_d}}(s) - V^{\pi_d, \nu_{est}}(s)| \leq \frac{\epsilon}{1-\gamma}\]

$$\sum_{t=T_i}^{T_{i+1}-1}(V^{*, \nu_{true}^{\pi_d}}(s_t)-V^{\pi_d, \nu_{true}^{\pi_d}}(s_t)) = \sum_{t=T_i}^{T_{i+1}-1}(V^{*, \nu_{true}^{\pi_d}}(s_t)-V^{\pi_d, \nu_{est}}(s_t) + V^{\pi_d, \nu_{est}}(s_t) - V^{\pi_d, \nu_{true}^{\pi_d}}(s_t))$$

$$\leq \sum_{t=T_i}^{T_{i+1}-1} |V^{*, \nu_{true}^{\pi_d}}(s_t)-V^{\pi_d, \nu_{est}}(s_t)| + \sum_{t=T_i}^{T_{i+1}-1} |V^{\pi_d, \nu_{est}}(s_t) - V^{\pi_d, \nu_{true}^{\pi_d}}(s_t)|$$

$$=\frac{\epsilon}{1-\gamma} (T_{i+1}-T_i) + \sum_{t=T_i}^{T_{i+1}-1} |V^{\pi_d, \nu_{est}}(s_t) - V^{\pi_d, \nu_{true}^{\pi_d}}(s_t)|$$

$V^{\pi_d, \nu_{est}}(s_t)$ denotes the expected cumulative discounted reward fetched by following $\pi_d$ using the estimated rewards while $V^{\pi_d, \nu_{true}^{\pi_d}}(s_t)$ denotes the expected cumulative discounted reward fetched by following $\pi_d$ using the true defender rewards. Since $\max_{s,a}|r_s^a-\tilde{r}_s^a|<\epsilon$, 

$$|V^{\pi_d, \nu_{est}}(s_t) - V^{\pi_d, \nu_{true}^{\pi_d}}(s_t)| \leq (1+\gamma+\gamma^2+...)\max_{s,a}|r_s^a-\tilde{r}_s^a| = \frac{\epsilon}{1-\gamma}$$ Combining the above inequalities, we have

$$\sum_{t=T_i}^{T_{i+1}-1}(V^{*, \nu_{true}^{\pi_d}}(s_t)-V^{\pi_d, \nu_{true}^{\pi_d}}(s_t)) \leq \frac{2\epsilon}{1-\gamma} (T_{i+1}-T_i)$$

Hence, the average regret in $[T_i, T_{i+1})$ is bounded as follows if the reward estimates in this interval satisfy $\max_{s,a}|r_s^a-\tilde{r}_s^a|<\epsilon$:

$$\frac{1}{T_{i+1}-T_i} \sum_{t=T_i}^{T_{i+1}-1}(V^{*, \nu_{true}^{\pi_d}}(s_t)-V^{\pi_d, \nu_{true}^{\pi_d}}(s_t)) = AvgRegret(T_i, T_{i+1}) \leq \frac{2\epsilon}{1-\gamma}$$

\section{Comparing MTD Approaches}

Game-theoretic methods leverage prior knowledge about attacker side payoffs, whereas other methods operate without such knowledge. 

The Markov Games model can capture the multi-stage nature of defender-attacker interactions. However, multi-armed bandit (MAB) based approaches lack the capability to model long-term temporal dependencies when selecting actions, unlike MDP-based approaches, which can capture the impact of current actions on future rewards and transitions.

Switching costs are dependent on the current configuration and switching action, a feature that can be incorporated into the payoff structure in game-theoretic settings. Conversely, simple MAB formulations struggle to model the dependency of switching costs on the current configuration. Employing a contextual MAB can address this limitation by integrating the current configuration into the context. However, it still suffers from a lack of modeling long-term dependencies, similar to other MAB methods. MDP-based approaches do not encounter this issue, as rewards inherently depend on the state and action. 

Finally, while effective real-time adaptation to evolving attack landscape is feasible in online bandit algorithms, it remains challenging in game-theoretic and MDP based models. However, our method ATA-FMDP can handle this non-stationarity in the environment.

\begin{table}[h]
\centering
\begin{tabular}{|l|c|c|c|c|c|}
\hline
 & Game-Theoretic & MAB & Contextual MAB & MDP & ATA-FMDP (ours) \\
\hline
No Attacker-side Payoffs Needed & $\times$ & $\checkmark$ & $\checkmark$ & $\checkmark$ & $\checkmark$ \\
Long Term Dependencies & $\checkmark$ & $\times$ & $\times$ & $\checkmark$ & $\checkmark$ \\
Incorporating Switching Costs & $\checkmark$ & $\times$ & $\checkmark$ & $\checkmark$ & $\checkmark$ \\
Tackling Adversary Evolution & $\times$ & $\checkmark$ & $\checkmark$ & $\times$ & $\checkmark$ \\
\hline
\end{tabular}
\vspace{1em}
\caption{Comparison of MTD Approaches}
\label{tab:mtc_comparison}
\end{table}

\section{Learning Properties of Unknown Attackers}
\label{appen:dh-unknown}
We consider an attack landscape in the web application domain containing a database hacker similar to the one considered in the main paper. However, this time, it can only hack $PostgreSQL$ and not both the databases like the $DH$ attacker from the main paper. This means that the defender should learn to stay in or switch to the $MySQL$ database to avoid the attacker altogether. However, we assume that this attack landscape is unknown to the defender here. We can see that the graphs do not depict much regarding the nature of the attacks except for the fact that the defender is able to achieve high values by learning from the interactions with the attacker, indicating that the attacker is not very strong and that $ATA-FMDP$ is able to learn the strategy of avoiding configurations with $PostgreSQL$ database to avoid it. However, if we look at the action factors and the number of switches in each factor, we will see that the database factor changes from $PostgreSQL$ to $MySQL$ $100\%$ of the time i.e., a switch is always made in the database factor while in the language factor, the two kinds of switches, i.e., switching to $Python$ and switching to $PHP$, are chosen equally. This helps us draw inferences regarding the attacks and shows that there might be a vulnerability in $PostgreSQL$ that the attacker is trying to exploit, and suitable actions can be taken to investigate it.\\

\begin{figure}[h]
     \centering
     \vspace{-1em}
     \begin{subfigure}[b]{0.3\textwidth}
         \centering
         \includegraphics[width=\textwidth, height=0.7\textwidth]{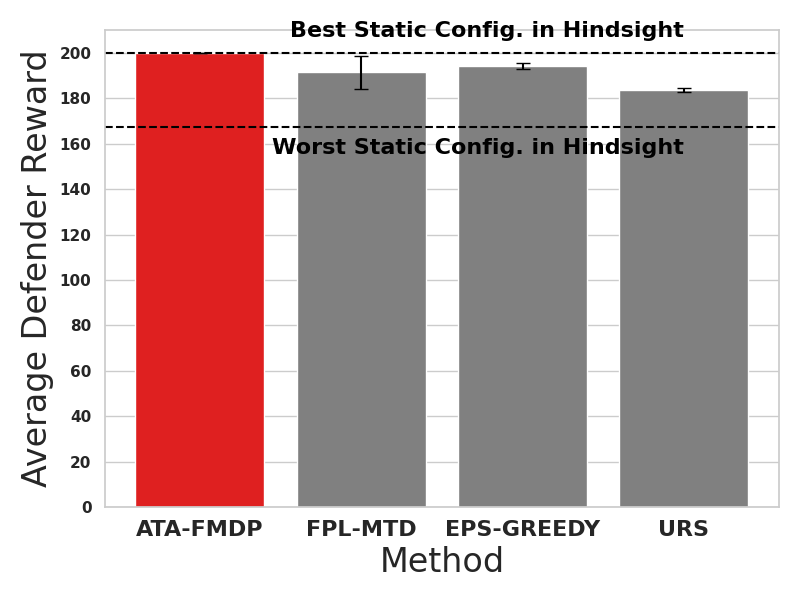}
         \caption{$\alpha = 0$}
     \end{subfigure}
     \hfill
     \begin{subfigure}[b]{0.3\textwidth}
         \centering
         \includegraphics[width=\textwidth, height=0.7\textwidth]{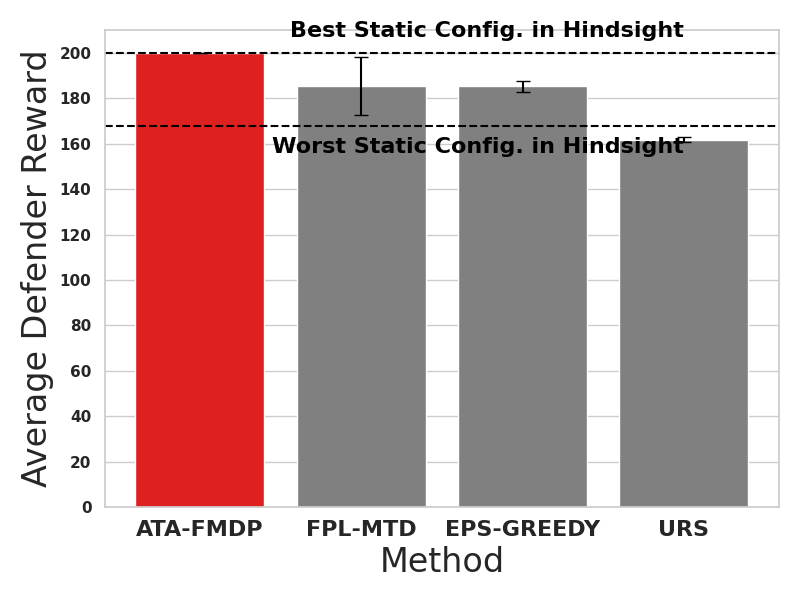}
         \caption{$\alpha = 0.5$}
     \end{subfigure}
     \hfill
     \begin{subfigure}[b]{0.3\textwidth}
         \centering
         \includegraphics[width=\textwidth, height=0.7\textwidth]{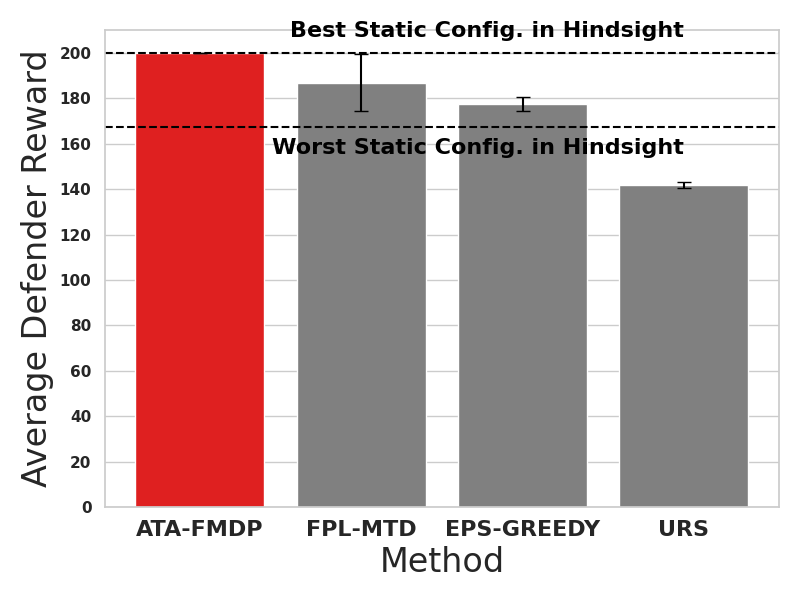}
         \caption{$\alpha = 1.0$}
     \end{subfigure}
     \hfill
      \begin{subfigure}{0.3\linewidth}
        \centering
        \includegraphics[width=\linewidth, height=0.6\textwidth]{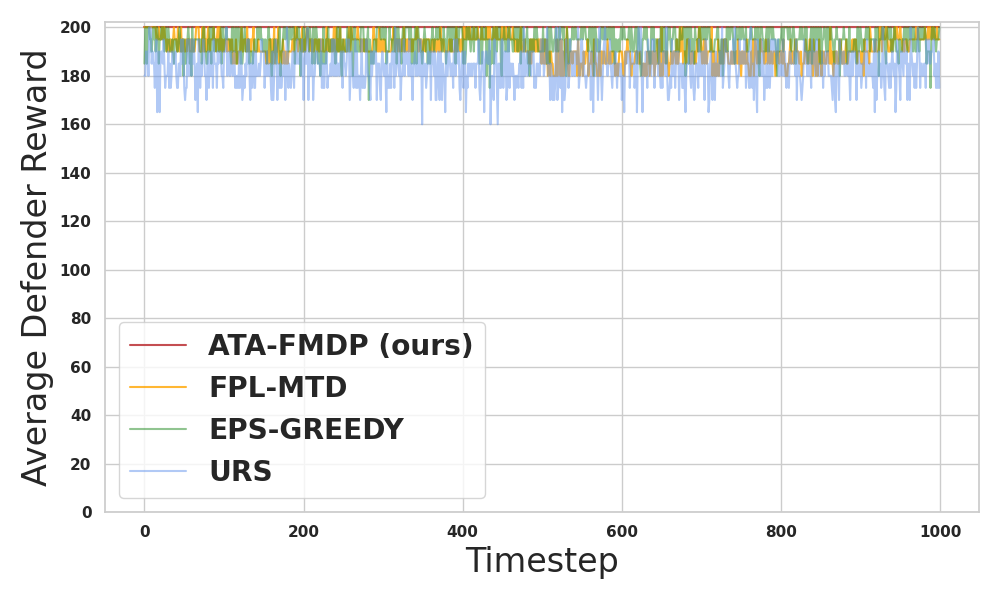}
        \caption{$\alpha = 0$}
      \end{subfigure}
      \hfill
      \begin{subfigure}{0.3\linewidth}
        \centering
        \includegraphics[width=\linewidth, height=0.6\textwidth]{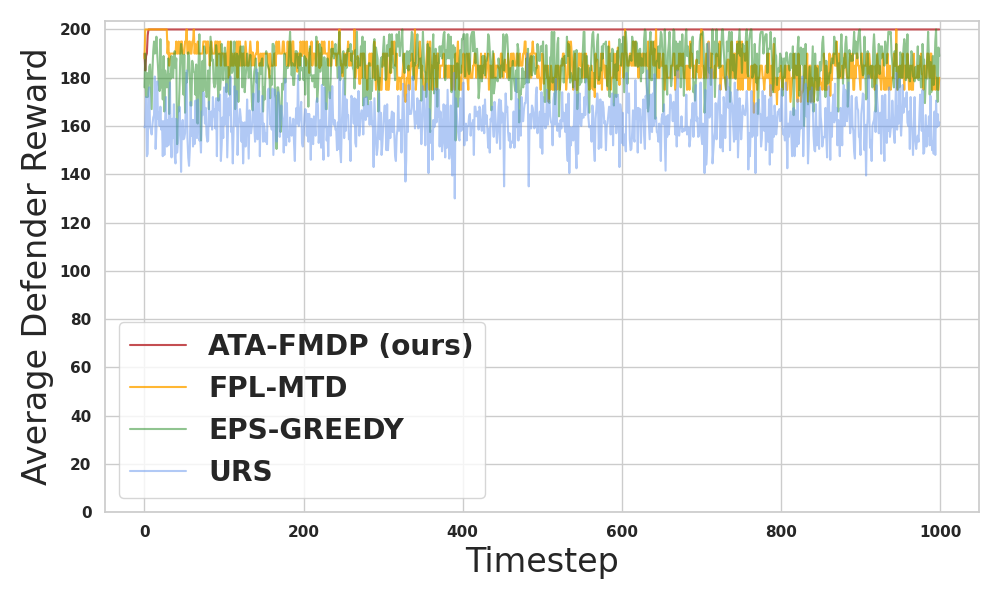}
        \caption{$\alpha = 0.5$}
      \end{subfigure}
      \hfill
      \begin{subfigure}{0.3\linewidth}
        \centering
        \includegraphics[width=\linewidth, height=0.6\textwidth]{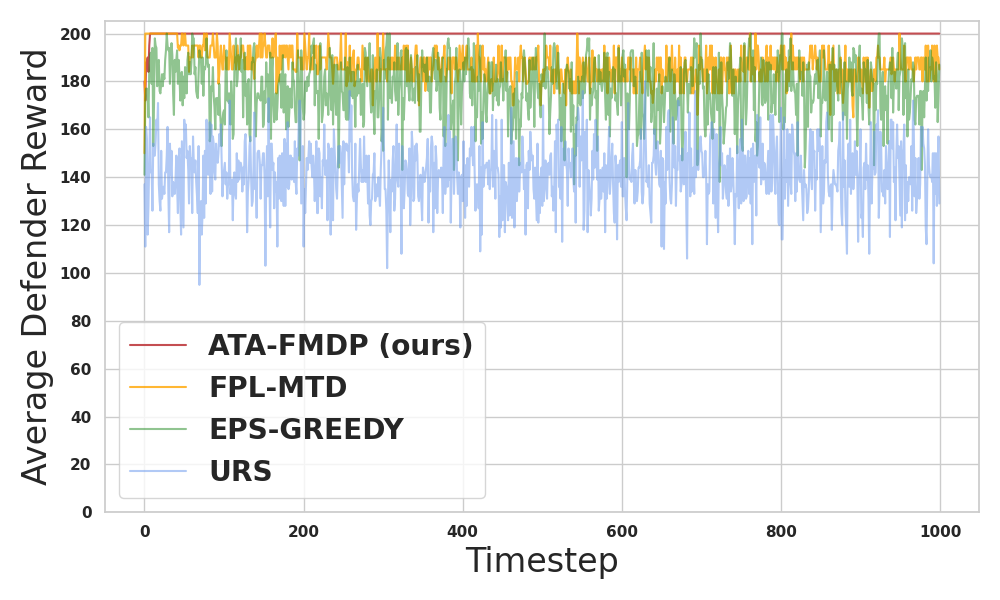}
        \caption{$\alpha = 1.0$}
      \end{subfigure}
      \vspace{1em}
      \caption{Attack Landscape of a PostgreSQL Database Hacker. The top three graphs show cumulative defender rewards for each defender strategy, while the bottom three graphs show the evolution of defender rewards over the 1000 timesteps. $\alpha$ refers to the relative weight given to switching costs.\\}
      \vspace{1em}
      \label{fig:appen-exp-2}
\end{figure}

\section{Effect of High Switching Cost}
\label{appen:high-sc}
We explore an adaptation of the evolving attacker landscape in the web application domain described in the paper, with switching costs increased threefold compared to the original setting. As anticipated, the adaptability of all the methods diminish in the face of greater switching costs. However, our approach demonstrates resilience in adapting to the evolving landscape of higher switching costs, while other methods struggle to do so. In particular, $URS$ experiences the most substantial decline in performance, given its inherent lack of sensitivity to switching costs. As a result, we observe that the average reward obtained from $URS$ falls below $30$ while other methods still maintain it above $100$.
\\

\begin{figure}[h]
     \centering
     \vspace{-1em}
     \begin{subfigure}[b]{0.4\textwidth}
         \centering
         \includegraphics[width=\textwidth, height=0.75\textwidth]{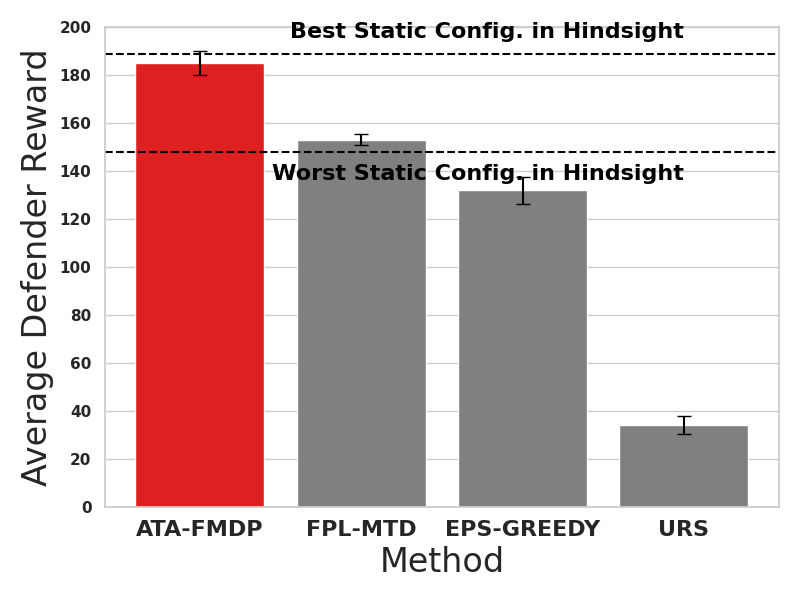}
         \caption{}
     \end{subfigure}
      \begin{subfigure}{0.4\linewidth}
        \centering
        \includegraphics[width=\linewidth, height=0.6\textwidth]{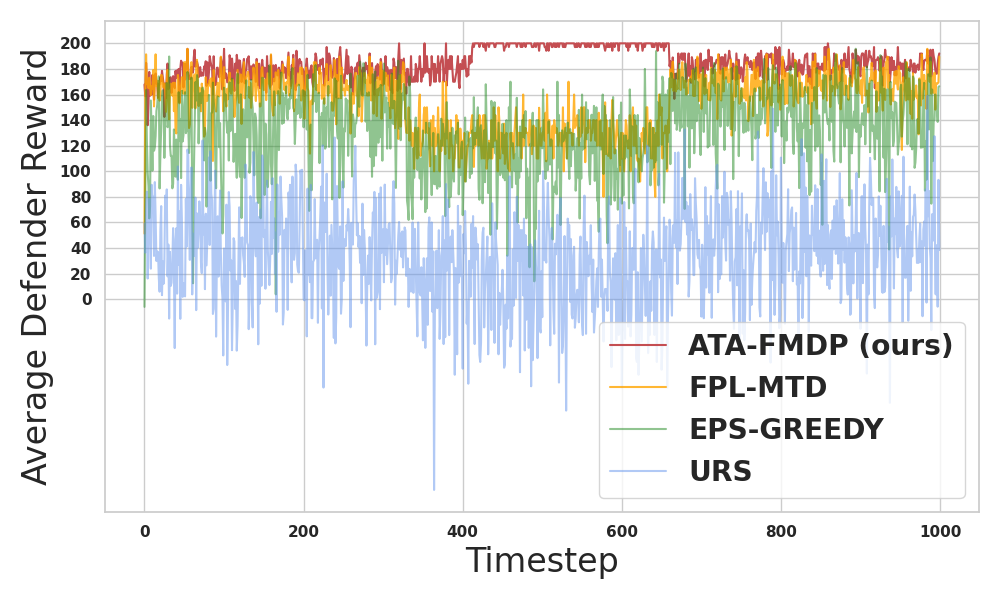}
        \caption{}
      \end{subfigure}
      \vspace{1em}
      \caption{Web Application Environement - Evolving Attack Landscape in with prevailing unknown type attacker between timesteps 330 and 660 but with switching costs increased threefold compared to the original setting.\\}
      \vspace{1em}
      \label{fig:appen-exp-3}
\end{figure}

When the same high switching cost setting is applied to the network environment, we see $FPL-MTD$ trying to progressively learn better strategies but is rather slow. On the other hand, $EPS-GREEDY$ shows better learning speed. $URS$ shows no learning as expected and remains at a poor reward due to high switching costs throughout the $T$ timesteps. Our approach, $ATA-FMDP$, outperforms other methods. Moreover, due to the high switching costs, it learns to avoid switching off any node when the unknown attacker prevails. This happens because in order to avoid the unknown attacker, the defender has to keep Node 0 offline, but the loss from keeping a node offline $(150)$ is far more than the loss from the unknown attack $(100)$. 
\\

\begin{figure}[h]
     \centering
     \vspace{-1em}
     \begin{subfigure}[b]{0.4\textwidth}
         \centering
         \includegraphics[width=\textwidth, height=0.7\textwidth]{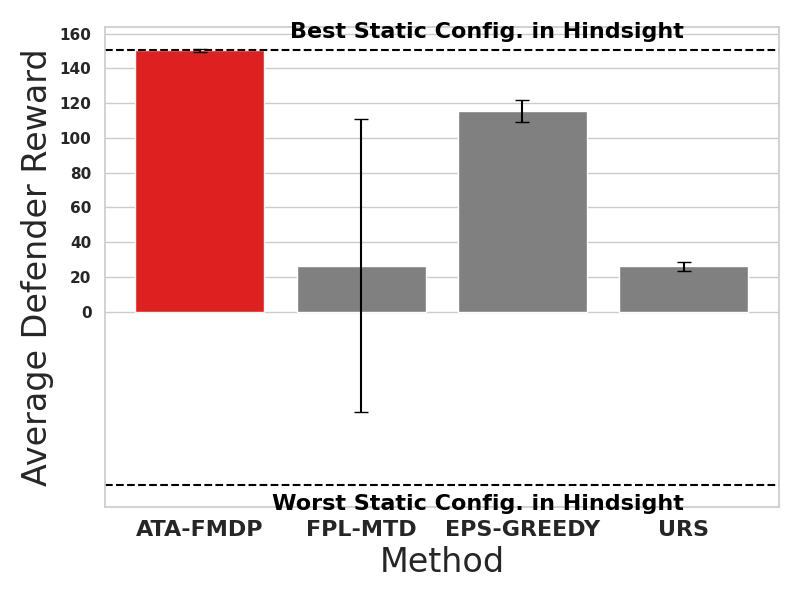}
         \caption{}
     \end{subfigure}
      \begin{subfigure}{0.4\linewidth}
        \centering
        \includegraphics[width=\linewidth, height=0.6\textwidth]{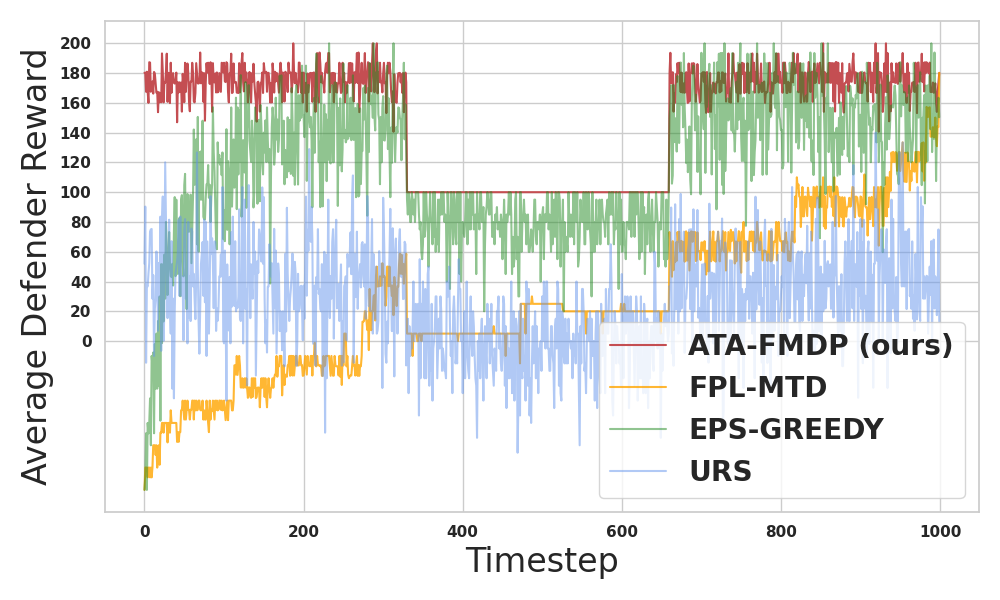}
        \caption{}
      \end{subfigure}
      \vspace{1em}
      \caption{Network Environment - Evolving Attack Landscape with prevailing unknown type attacker between timesteps 330 and 660 but with switching costs increased threefold compared to the original setting.\\}
      
      \label{fig:appen-exp-4}
\end{figure}

\section{Domain Diagrams}
\label{appen:domain-diag}
\begin{figure}[h]
     \centering
     \vspace{-1em}
         \includegraphics[width=0.65\textwidth, height=0.35\textwidth]{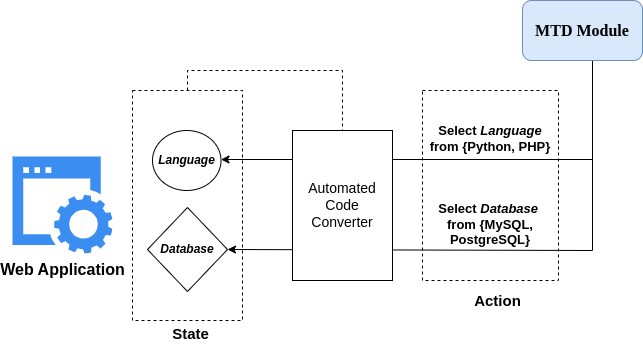}
      \vspace{1em}
      \caption{MTD in Web Application\\}
      \vspace{2em}
      \label{appen:webapp-domain}
\end{figure}

\begin{figure}[h]
     \centering
     \vspace{-1em}
         \includegraphics[width=0.55\textwidth, height=0.5\textwidth]{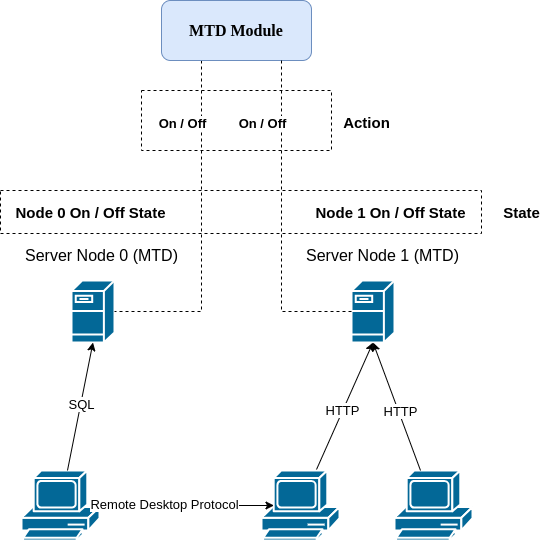}
      \vspace{1em}
      \caption{MTD in Network\\}
      \vspace{2em}
      \label{appen:network-domain}
\end{figure}

Figures \ref{appen:webapp-domain} and \ref{appen:network-domain} depict the Web Application environment and a possible instance of the Network Environment, respectively.
\vspace{1em}

\section{Reproducibility}
To ensure the reproducibility of the code, a seed value of $10$ has been used to initialize a pseudo-random number generator in all our programs. All the datasets used, as well as the values of the hyperparameters, have been included in the code; these values have also been presented in Table \ref{tab:hyper-params}.

\begin{table}[h]
\centering
\begin{tabular}{|c | c|} 
 \hline
 Parameter & Value \\
 \hline
    seed & 10\\
    no. of iterations & 10\\
    no. of timesteps per iteration (T) & 1000\\
    Discount factor ($\gamma_{}$) & 0.9\\
    M & 200.0 \\
    $\beta$ & 2\\
    $\alpha$ & $[ 0.0, 0.5, 1.0 ]$ \\
    $\gamma_{FPL-MTD}$ & 0.007\\
    $\eta_{FPL-MTD}$ & 0.1\\
    $Lmax_{FPL-MTD}$ & 1000\\ 
    $\epsilon_{EPS-GREEDY}$ & 0.2 \\ [1ex] 
 \hline
\end{tabular}
\caption{Parameter Values}
\label{tab:hyper-params}
\end{table}

\end{document}